\documentclass[aps,prb,a4paper,twocolumn,showpacs,showkeys,floatfix, superscriptaddress]{revtex4-2}

\usepackage{graphicx}%for pdf-figures
\usepackage{color}
\usepackage{amsmath}
\usepackage{amssymb}
\usepackage{multirow}
\usepackage{hyperref}

\newcommand{\nabok}[2]{#1Cu$_7$(TeO$_4$)(SO$_4$)$_5$#2}

\newcommand{\abs}[1]{\left|#1\right|}
\renewcommand{\vec}[1]{\mathbf{#1}}
\begin{document}

\title{Collinear and noncollinear antiferromagnetic ordering in a highly frustrated decorated square kagom\'{e} lattice antiferromagnets of the nabokoite family}

\author{V. N. Glazkov}
\email{glazkov@kapitza.ras.ru}
\affiliation{ P.L. Kapitza Institute for Physical Problems, RAS, Kosygina 2, Moscow 119334, Russia}

\author{Ya. V. Rebrov}
\affiliation{ P.L. Kapitza Institute for Physical Problems, RAS, Kosygina 2, Moscow 119334, Russia}

\author{M.~A.~Dubovitskii}
\affiliation{ P.L. Kapitza Institute for Physical Problems, RAS, Kosygina 2, Moscow 119334, Russia}
\affiliation{National Research University  ``HSE'', 109028, Moscow, Russia}

\author{M.~M.~Markina}
\affiliation{M.V.~Lomonosov Moscow State University, Moscow 119991, Russia}
\affiliation{National University of Science and Technology ``MISiS'', Moscow 119049, Russia}

\author{K.V.~Zakharov}
\affiliation{M.V.~Lomonosov Moscow State University, Moscow 119991, Russia}

\author{A.~F.~Murtazoev}
\affiliation{M.V.~Lomonosov Moscow State University, Moscow 119991, Russia}

\author{P.~S.~Berdonosov}
\affiliation{M.V.~Lomonosov Moscow State University, Moscow 119991, Russia}
%\affiliation{National University of Science and Technology ``MISiS'', Moscow 119049, Russia}

\author{A.~N.~Vasiliev}
\affiliation{M.V.~Lomonosov Moscow State University, Moscow 119991, Russia}
\affiliation{National University of Science and Technology ``MISiS'', Moscow 119049, Russia}

\begin{abstract}
Nabokoite family compounds \nabok{A}{X}{} (A=Na, K, Rb, Cs; X=Cl, Br) host frustrated 2D square kagom\'{e} lattice layers decorated by additional inter-layer magnetic ions. Despite the high Curie-Weiss temperatures $\Theta\simeq 150$~K all nabokoite family compounds order only at $T_\textrm{c}\simeq3-6\textrm{ K}\ll\Theta$ highlighting role of magnetic frustration in nabokoites. We study magnetic order in nabokoites with multi-frequency electron spin resonance spectroscopy and thermodynamic measurement (specific heat, magnetization and dielectric permittivity). Our study reveals that the choice of the low-temperature ground state is qualitatively different in light-alkali-ion (A=K, Na) and heavy-alkali-ion (A=Rb, Cs) compounds. Heavy-alkali-ion nabokoites order in conventional collinear antiferromagnetic pattern with easy-axis anisotropy. The  parameters of the ordered antiferromagnetic state are very close for all heavy-alkali-ion subfamily: zero-field magnon gap at 1.7~K is about 120~GHz and spin-flop-field is about 3.5~T. Light-alkali-ion members of nabokoite family demonstrate much more complicated route to the ordered state: firstly, a ferroelectric transition at 25-90~K partially lifts the frustration and thus pre-cooks the low-temperature ordering; secondly, an unusual noncollinear magnetic order develops via two-step phase transition with first transition temperature $T_\textrm{c1}\simeq 5-6$~K and the second transition at $T_\textrm{c2}\simeq 3-4$~K. Noncollinear order is evidenced by observation of characteristic non-Larmor antiferromagnetic resonance mode. Spin dynamics of light-alkali-ion nabokoites is characterized by three zero-field magnon gaps and two spin-reorientation fields and, contrary to the heavy-alkali-ion subfamily, the values of magnon gaps and critical fields are quite different for different light-alkali-ion compounds. The finite-size cluster modeling of pyramidal structural block of nabokoite structure combined suggests that  the critical closeness of the nabokoite exchange coupling parameters to the border-line between the different quantum ground state of pyramidal building block of nabokoite structure could be the clue to the choice of qualitatively different ordered state in light- and heavy-alkali-ion nabokoites.
\end{abstract}

\date{\today}
\keywords{two-dimensional magnet, magnetic frustration, antiferromagnetic order}

\pacs{75.50.Ee, 76.50.+g, 75.25.+z}

%75.50.Ee Antiferromagnetics
% 76.50.+g Ferromagnetic, antiferromagnetic, and ferrimagnetic resonances; spin-wave resonance
% 75.25.+z Spin arrangements in magnetically ordered materials
\maketitle

\section{Introduction}
Geometrically frustrated magnets remains a major focus of magnetism during the recent decades \cite{lacroix, ramirez,starykh,ramirez2025}. Specific architecture of the exchange bonds results in a fine balance of the strong exchange coupling effects which  prevents conventional magnetic ordering at $T \simeq \Theta_\textrm{CW}$. As a result, frustrated system either remains  in a paramagnetic spin-liquid state down to $T=0$ or orders at $T\ll  \Theta_\textrm{CW}$ under the influence of additional, usually much weaker, interactions or via the ``order-by-disorder'' mechanism \cite{order-disorder1,order-disorder2}. Since the final choice of the ordered state results from the interplay of  simultaneously acting low-energy-scale interactions, the magnetic phases of the frustrated magnets sometimes turn out to be highly unusual, fluctuations-stabilized collinear phase in the triangular lattice antiferromagnet \cite{chubukov,smirnov,kenzelman} or a complicated partially ordered state of Heisenberg pyrochlore magnet Gd$_2$Ti$_2$O$_7$ \cite{GTO1,GTO2} being the well known examples. The change of temperature, magnetic field, ambient or chemical pressure affects interactions balance yielding rich phase diagrams \cite{starykh,kenzelman,petrenko,GTO-torque} which are of interest. Lattice degrees of freedom can be also important for the frustrated magnets as minute lattice modification can violate fable balance of interactions within the frustrated magnet lifting the frustration and leading to the formation of magnetic order via spin-Peierls-like mechanism \cite{lacroix,Zn-spinel,spinel2}.

Geometric frustration usually appears when a triangular building blocks exist within the exchange bonds network. This makes 2D triangular lattice and a kagom\'{e} lattice a long time favorite models of frustrated magnetism. Square kagom\'{e} lattice (SKL) is another example of corner-sharing triangles network. The ideal SKL consists of equilateral triangles grouped around alternating square and octagonal voids and is predicted to be a gapped spin-liquid \cite{SidGeorge,Richter-skl-gapped}. However, extensions of this model including inequivalent couplings along the triangle sides and next-nearest-neighbor interactions demonstrate variety of the possible ordering patterns \cite{Lugan,morita-j123,Gembe-noncomplanar}.

Nabokoite family compounds \nabok{A}{X}{} (here A=Na, K, Rb, Cs and X=Cl, Br), referred below as A/X-nabokoites for short,  host frustrated two-dimensional square  kagom\'{e} lattice (SKL) which is additionally decorated by inter-layer copper ions \cite{alisher}.  Magnetic susceptibility measurements suggest antiferromagnetic Curie-Weiss temperatures $\Theta_\textrm{CW}\simeq 100...150$~K\cite{alisher,markina,china}, indicating presence of strong antiferromagnetic bonds. All nabokoite family magnets remain paramagnetic well below the Curie-Weiss temperature, indicating high degree of geometric frustration, and order antiferromagnetically at 3...6~K \cite{alisher,markina}. Study of the magnetic order in nabokoites provides an insight on the role of the minute details of inter-layer interactions on the choice of the magnetically ordered state in frustrated magnet.

Our electron paramagnetic resonance (EPR) study in nabokoite family magnets reported separately \cite{nabok-pm} demonstrated by virtue of accurate absolute calibration of EPR absorption presence of the  decoupled magnetic subsystems  in nabokoite lattice: only a fraction of copper spins in nabokoite lattice contributes to the paramagnetic absorption and the same fraction orders below the N\'{e}el point. Earlier studies \cite{markina} demonstrated that  the antiferromagnetic resonance spectra in K/Cl nabokoite are typical for a noncollinear antiferromagnet. A study of the dynamic and static dielectric properties  revealed that the light-alkali-ion Na/Cl and K/Cl nabokoites undergo structural transitions at approximately 90~K (Na/Cl) and 25~K (K/Cl) marked by $\varepsilon'$ and $\varepsilon''$ anomalies, while the heavy-alkali-ion Cs/Cl and Rb/Cl nabokoites demonstrate no such features \cite{markina,china,rebrov}.

In the present research we report detailed antiferromagnetic resonance (AFMR) study of the ordered phases of nabokoite family compounds. We have observed formation of noncollinear ordering for all light-alkali-ion compounds (K/Cl, Na/Cl and K/Br -- nabokoites) and conventional collinear easy-axis antiferromagnetic ordering for heavy-alkali-ion compounds (Cs/Cl, Rb/Cl, Cs/Br and Rb/Br). Characteristic features of the AFMR spectra allowed to determine values of magnon gaps and spin-reorientation transitions fields. Temperature evolution of the magnon gaps clearly demonstrates that the noncollinear order is formed by a two-stage phase transition, while collinear ordering develops in a usual single-stage phase transition.  Complimentary  dielectric properties check-up of bromine nabokoites revealed presence of high-temperature dielectric anomaly in K/Br compound at 60~K and absence of similar dielectric anomalies in Cs/Br and Rb/Br nabokoites. The choice of the ordering pattern in nabokoites turns out to be  pre-determined by high-temperature dielectric transition. Partial ordering of the nabokoites spins is discussed compared to the model of decorated SKL \cite{alisher,markina} and DFT-suggested exchange network \cite{DFT}.

\section{Samples and experimental techniques}

\subsection{Samples}

\begin{figure}[th]
\centering
  \includegraphics[width=\columnwidth]{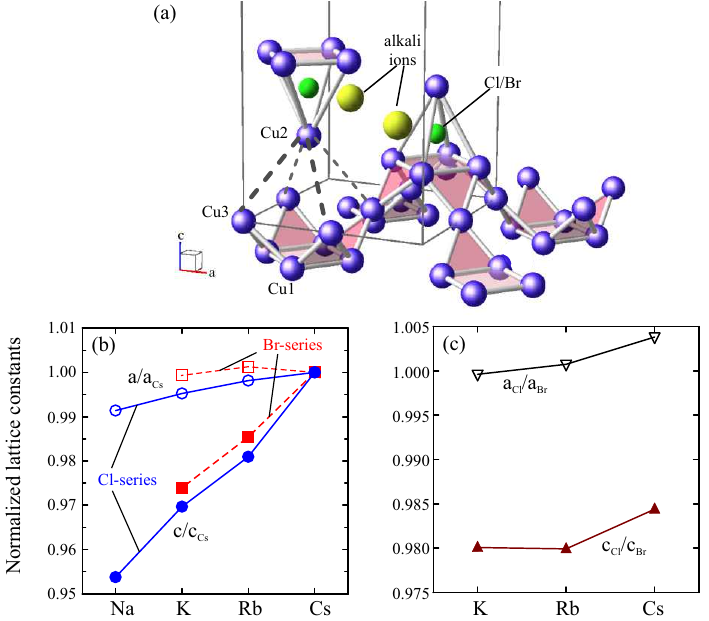}\\
  \caption{(color online) (a) Fragment of nabokoite crystal structure. Only copper, halogen and alkali ions positions are shown.   (b) Nabokoites lattice constants normed to the corresponding constants for  Cs/Cl or Cs/Br-compounds. Open symbols --- $a/a_\textrm{Cs}$, closed symbols --- $c/c_\textrm{Cs}$. Circles (solid lines) --- chlorine series of compounds, squares (dashed lines) --- bromine series of compounds. (c) Ratios of the lattice constants for the chlorine and bromine nabokoite series. Open symbols --- $a$, closed symbols $c$.}
  \label{fig:struct}
\end{figure}

Polycrystalline samples of nabokoite family compounds \nabok{A}{X}{} (here A=K, Na, Cs, Rb and X=Cl, Br) were synthesized as described in Ref.~\cite{alisher}. Seven members of this family save for the Na/Br nabokoite were obtained. Samples were the same as the samples  used in our earlier studies \cite{markina,rebrov}.

All nabokoite family compounds are iso-structural with tetragonal P4/ncc (D$_\textrm{4h}^8$) space group at room temperature. X-ray structural analysis excludes presence of other crystallographic phases in measurable quantities. Copper ions occupy three non-equivalent positions (Fig.~\ref{fig:struct}-a): six of the seven copper ions per nabokoite molecule  form buckled 2D square kagom\'{e} layers (Cu1 and Cu3 positions), while seventh ion  takes interlayer position (Cu2). Halogen and alkali ions are at inter-layer positions: halogen (Cl/Br) ion takes  symmetric position at the axis of Cu1-Cu2 square-based pyramid, alkali ions are located aside from the copper-copper bridging bonds almost on the same level along the tetragonal $c$-axis as the decorating inter-layer Cu2 ions. Change  of  halogen or alkali ion affects mostly the inter-layer spacing between the SKL layers \cite{alisher,alisher-phd} (Fig.~\ref{fig:struct}-b,c): relative change of the  lattice constant $c$ is  approximately five-fold compared to the relative change of the in-plane lattice constant $a$.

\subsection{Electron spin resonance}
Electron spin resonance (ESR) measurements were performed using a set of  home-made transmission-type microwave spectrometers covering microwave frequencies range 9--150~GHz equipped with helium-4 bath cryostat and 8~T superconducting magnet.  Nabokoite powder samples for ESR experiments were prepared by soaking 50-100~mg of nabokoite powder with the ethanol-diluted glue (``BF-2'' trademark, analogous to GE varnish) in a thin-wall paper container (container diameter $\approx 5$~mm, resulting sample height $\approx 1$~mm). Note, that at 30~GHz microwave wavelength equals 1~cm. Therefore, at frequencies below 15~GHz sample can be considered as a point sample subjected to the uniform microwave field, while above 50~GHz sample size  is comparable to the microwaves wavelength and microwave field within the sample includes mixture of different polarizations of high-frequency magnetic field and certain admixture of microwave electric field.

ESR is a sensitive tool to study magnetic phase transitions and to characterize ordered phases. Paramagnetic resonance above the N\'{e}el point  can be basically understood as the atomic spin flips upon absorption of a microwave quantum.  Antiferromagnetic resonance (AFMR) below the N\'{e}el point corresponds to the creation of collective excitations of antiferromagnet: absorption of the microwave quantum give birth to a $q=0$ spin-wave quantum (antiferromagnetic magnon). I.e., antiferromagnetic resonance experiment measures magnon energies in the Brillouin-zone center.

Spin waves eigenfrequencies are sensitive to the  type of magnetic order, crystal symmetry and applied magnetic field. Spin-reorientation phase transitions are usually accompanied by softening of a certain spin-wave mode. Detailed theoretical considerations of the antiferromagnetic resonance in collinear and noncollinear antiferromagnets are given in Refs.~\cite{kubo,gurevich,andmar,svistfar,noncolnum}. Here we will briefly recall the main results relevant for our analysis.

In the absence of anisotropic spin-spin interactions continuous symmetry of Heisenberg exchange interaction implies gapless spin-waves spectra in agreement with the Goldstone theorem. Quantity of the Goldstone modes depends on the type of magnetic ordering: two  modes in the case of collinear magnetic ordering (e.g., conventional two-sublattices magnetic ordering \cite{kubo,gurevich}), and three modes in the case of noncollinear magnetic ordering \cite{andmar,prozmar,glazkovgarnet}. Besides of the Goldstone modes, high-energy modes (also called exchange modes) could exist in many-sublattices antiferromagnets \cite{baryakhtar}. Characteristic feature of AFMR in noncollinear magnets is the ``anomalous'' asymptotic behavior of the resonance modes field dependences: while in the case of the collinear antiferromagnet field dependent resonance modes approaches Larmor frequency $\omega_\textrm{L}=\gamma B$ ($\gamma$ is a gyromagnetic ratio),  in the noncollinear case one of the field-dependent resonance modes has a non-Larmor behavior with the slope determined be the ratio  of magnetic susceptibilities of noncollinear magnetic structure \cite{svistfar,prozmar,zaliznyak}  (see also Appendix~\ref{sec:app-noncol}).

Anisotropic spin-spin interactions break continuous symmetry and Goldstone modes acquire gaps.  These gaps are proportional to the antiferromagnetic order parameter (sublattice magnetization), see e.g. \cite{kubo}, and develops from zero below the N\'{e}el temperature. Combination of these anisotropic interactions with the effects of applied magnetic field yield characteristic dependences of the spin oscillations eigenfrequencies on magnetic field, which can be conveniently presented as a frequency-field diagram $f(B)$. Analysis of $f(B)$ data provides information on the type of magnetic ordering (collinear or noncollinear), yields values of the  spin-wave gaps  and magnetic phase transition fields.

In the case of powder sample observed absorption spectra are additionally averaged over the powder particle orientations. This averaging results in formation of absorption bands with  edges of absorption  determined by the limiting particle orientations and absorption maxima determined by the statistical weight of antiferromagnetic eigenfrequencies distribution over the $f(B)$ plane. The AFMR absorption averaging for the collinear antiferromagnet is discussed in Appendix~\ref{sec:app-pow}.

Temperature dependence of the spin-wave gaps noted above can be exploited in a ``temperature resonance'' experiment by registering microwave power transmitted through the cavity with the sample as a function of temperature. Once,  at some temperature $T_0$, the spin-wave gap value  $\Delta(T_0)$ equals the energy of microwave quanta $2\pi\hbar f_0$ absorption of the microwave energy appears.  By repeating this experiment at different eigenfrequencies of multi-modal microwave cavity we can accurately collect data set $\lbrace T_0, f_0\rbrace$, which is nothing else but the temperature dependence of the spin-wave gap, the later being proportional to antiferromagnetic order parameter (sublattice magnetization).

\subsection{Dielectric properties measurements}
Temperature dependences of dielectric constant $\varepsilon$  have been measured at various frequencies from 1 to 20~kHz by a capacitance bridge Andeen-Hagerling 2700A on a thin pressed nabokoite pellets covered by a silver paste. No significant frequency dependence of dielectric permittivity was observed at low frequencies.

High-frequency ($\simeq 10$~GHz) dielectric losses were studied as  described in details in \cite{rebrov}. For this purpose small (5--20~mg) nabokoite powder sample was placed to the geometrical center of the rectangular microwave cavity (cavity dimensions $\simeq 7\times17\times 40$~mm$^3$). The sample at this location is subject either to microwave electric or microwave magnetic field depending on the choice of cavity resonance mode \footnote{Microwave TE10$n$ mode in rectangular cavity corresponds to the following pattern of microwave field distribution within the cavity: microwave field along the shortest side of the cavity is uniform, there is a single half-wave of electromagnetic field along the middle side of the cavity and there are $n$ half-waves along the longest side of the cavity. TE101 is the lowest frequency eigenmode of the rectangular cavity. Tangential component of microwave electric field is zero at cavity walls. Even $n$ case corresponds to the node of microwave electric field at the geometrical center of the cavity,  odd $n$ case corresponds to the peak of the microwave electric field at the cavity center. More details can be found in electrodynamics textbooks, e.g. \cite{jack} }. There is a node of  microwave electric field at the cavity center for TE102 mode (11.4~GHz for our cavity) and the node of  microwave magnetic field for TE101 and TE103 modes (9.4 and 14.1 GHz). With this sample location we can perform both standard electron spin resonance experiment by taking dependence of the microwave losses as a function of magnetic field (using TE102 mode), or can measure magnetic or dielectric microwave losses as a function of temperature (using TE102 or TE101/TE103 modes, correspondingly). High-frequency dielectric losses were sought for from 1.7~K to 200~K to check both for the dielectric absorption at the magnetic phase transition and for the high-temperature dielectric anomalies.

\subsection{Magnetization and specific heat}

Magnetization $M(B)$, magnetic susceptibility $\chi(T)$ and specific heat $C_\textrm{P}(T,B)$ curves were measured using the Quantum Design PPMS system. Powder sample was prepared by pressing the appropriate amount of \nabok{A}{X}{} powder within the plastic sample-holder for the  magnetic measurements or by preparing small pellet for the specific heat measurements. Magnetic susceptibility curves for all samples were measured both in field-cooling (FC) and zero-field-cooling (ZFC) regimes with no observable difference found.

\section{Experimental results}
\begin{table*}[th]
\caption{Phase transitions temperatures  for nabokoite family compounds \nabok{A}{X}{} (marked as A/X for short) from different experimental techniques: SH -- specific heat measurements, MM -- magnetic susceptibility measurements, ESR -- electron spin resonance, NMR -- nuclear magnetic resonance,  DP -- low-frequency (up to MHz range) dielectric properties measurements, HFDP -- high-frequency (GHz range) dielectric properties measurements. Reference [P.W.] stays for present work data.}
\label{tab:temperatures}
\begin{ruledtabular}
\begin{tabular}{cccc}
&\multicolumn{2}{c}{T$_\textrm{c}$, K and Reference}& Dielectric absorption\\
Compound& Magnetic&High-temperature dielectric &close to the magnetic \\
& phase transition&phase transition&transition\\
\hline
Na/Cl&$3.7\pm0.2$; $5.8\pm0.2$ (ESR [P.W.])&$90\pm10$ (HFDP \cite{rebrov})&observed (HFDP \cite{rebrov})\\
&$3.7\pm0.1$; $5.7\pm0.2$ (SH [P.W], \cite {alisher})&&\\
\hline
K/Cl&3.2 (MM \cite{markina})&25.4 (DP \cite{markina})&observed (DP \cite{china})\\
&3.2; 5.7 (SH \cite{markina})&25 (HFDP \cite{rebrov})&\\
&4; 5.6 (MM \cite{china})&27; 30 (DP \cite{china})   &\\
&4.5 (NMR \cite{china})&&\\
&4.5; 5.5 (SH \cite{china})&&\\
&$3.2\pm0.2$ (ESR \cite{markina})&&\\
\hline
K/Br&$3.2\pm0.3$; $6.0\pm0.2$ (SH [P.W.])&$60\pm 2$ (DP [P.W.])&no data \\

\hline
Rb/Cl&$5.10\pm0.10$ (ESR [P.W.])&none \cite{rebrov}&observed (HFDP \cite{rebrov})\\
&4.8 (MM \cite{alisher})&&\\
&$5.5\pm0.2$ (SH [P.W.],\cite{alisher})&&\\
\hline
Rb/Br&$5.40\pm0.10$ (ESR [P.W])&none [P.W.]&observed (HFDP [P.W.])\\
&$5.45\pm0.12$ (SH [P.W.])&&\\
\hline
Cs/Cl&$5.10\pm0.10$ (ESR [P.W.])&none \cite{rebrov}&observed (HFDP \cite{rebrov})\\
&$6.0\pm0.4$    (SH [P.W.],\cite{alisher})&&\\
\hline
Cs/Br&$5.25\pm0.10$ (ESR [P.W.])&none [P.W.]&observed (HFDP [P.W.])\\
&$5.2\pm0.2$  (SH [P.W.],\cite{alisher})&&\\
\end{tabular}
\end{ruledtabular}
\end{table*}

\subsection{Low-temperature magnetic phase transitions in nabokoites}
\begin{figure}[th]
\centering
  \includegraphics[width=\columnwidth]{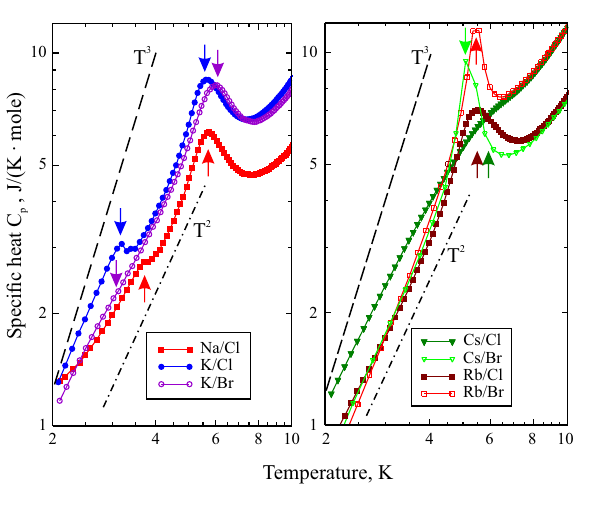}\\
  \caption{(color online) Low-temperature specific heat curves $C_\textrm{p}(T)$ for nabokoite family compounds \nabok{A}{X} (A=Na, K, Cs, Rb; X=Cl, Br). Left panel: light-alkali-ion nabokoites (A=Na, K). Right panel: heavy-alkali-ion nabokoites (A=Rb, Cs). Vertical arrows marks positions of specific heat peaks corresponding to the low-temperature phase transitions (see Table~\ref{tab:temperatures}). Dashed and dash-dotted lines correspond to $T^3$ and $T^2$ dependences expected for 3D and 2D Heisenberg antiferromagnets, correspondingly.}\label{fig:specheat}
\end{figure}

All  nabokoite family compounds \nabok{A}{X}{}  (A=Na, K, Rb, Cs; X=Cl, Br) order antiferromagnetically at low temperatures.  Magnetization and specific heat measurements for powdered chlorine compounds were reported in \cite{alisher,markina}, dielectric properties of the chlorine series were studied in \cite{rebrov,markina,china}, results of variety of experimental techniques applied to single-crystalline K/Cl compound were reported in \cite{china}.

Low temperature specific heat curves for all nabokoites samples are gathered at Fig.~\ref{fig:specheat}. All samples demonstrate peak features related to low-temperature phase transitions. Quality of the specific heat peaks is sample-dependent, which is probably related to the sample quality issues. The low-temperature specific heat behavior is distinct for the light-alkali-ion nabokoites (A=Na, K) and heavy-alkali-ion nabokoites (A=Rb, Cs):  all light-alkali-ion compounds demonstrate two low-temperature peaks at $T_\textrm{c1}\approx 5...6$~K and at $T_\textrm{c2}\approx3...4$~K (low-temperature peak for K/Br compound is strongly smeared), while all heavy-alkali-ion nabokoites demonstrate single peak at approximately 5.5~K (peaks for chlorine compounds are more smeared).

Earlier research \cite{markina} interpreted higher-temperature $T_\textrm{c1}$ peak in K/Cl nabokoite as a possible trace of a singlet gap expected for the ideal SKL magnet \cite{Richter-skl-gapped,tomczak}. We will demonstrate in one of  the following subsection that the magnon gap opens at this temperature, which is a direct evidence of a magnetic phase transition. This is also supported by single-crystalline measurements of Ref.~\cite{china}: specific heat peak at approximately 6~K became sharp  in single-crystalline K/Cl nabokoite sample, as expected for the phase transition.

Magnetic ordering in nabokoites is accompanied by a weak anomaly in dielectric properties pointing to the possible multiferroicity of these materials.  High-frequency ($\simeq 10$~GHz) dielectric properties study \cite{rebrov} revealed weak dielectric absorption in the vicinity of the phase transition for chlorine nabokoites (save for the K/Cl compound where low-temperature dielectric measurements were hindered by the closely located ferroelectric anomaly). Weak $\varepsilon'(T)$ peak was recently observed at $T\approx 5$~K in the low-frequency (1~kHz--1~MHz) experiments on single-crystalline K/Cl nabokoite   \cite{china}. Similar dielectric absorption anomalies were observed in Rb/Br and Cs/Br nabokoites over the course of the present research.

Additionally, high-temperature (25...90~K) dielectric anomalies were found in K/Cl and Na/Cl compounds \cite{markina,rebrov,china} while no such anomalies were observed in Cs/Cl and Rb/Cl nabokoites. Single-crystalline experiments for  K/Cl compound \cite{china} revealed  that high-temperature dielectric anomaly is related to the ferroelectric ordering  with electric polarization reaching value $P \simeq 30~\mu\textrm{C}/\textrm{m}^2$ at 5~K. We assume, that similar high-temperature  dielectric anomalies observed in other light-alkali-ion nabokoites are also due to the ferroelectric ordering.

Positions of low temperature phase transitions determined by different experimental techniques and positions of high-temperature dielectric anomalies  (including the data described in the following subsections of the present paper) are summarized in Table~\ref{tab:temperatures}. Difference in transition temperature values can be reasonably ascribed to the difference in sample quality and to the interpretation of the particular experimental technique results.

Table \ref{tab:temperatures} highlights systematic difference of the light- and heavy-alkali-ion nabokoites: the light-alkali-ion compounds demonstrate high-temperature dielectric anomaly and two magnetic phase transitions at low temperatures, while the heavy-alkali-ion nabokoites demonstrate no high-temperature anomalies and a single magnetic phase transition.

\subsection{Dielectric anomalies in \nabok{A}{Br} (A=K, Rb, Cs)}
\begin{figure}[th]
\centering
  \includegraphics[width=\columnwidth]{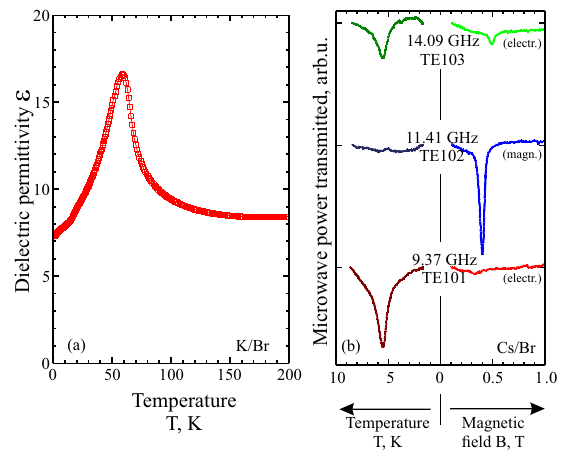}\\
  \caption{(color online)  (a) Temperature dependence of the dielectric permittivity $\varepsilon$ for \nabok{K}{Br}. Experiment frequency $f=10$~kHz. (b) Magnetic resonance at $T=1.7$~K (right horizontal semi-axis) and ``temperature resonance'' at $B=0$ (left horizontal semi-axis) demonstrating low-temperature dielectric absorption at magnetic phase transition  for \nabok{Cs}{Br}{}. Sample is placed in the geometrical center of the rectangular microwave cavity, ``(electr.)'' and ``(magn.)'' notes mark experiments with the sample positioned in the maximum of microwave electric and magnetic field correspondingly.}\label{fig:dielectric}
\end{figure}

High-frequency (GHz range) dielectric behavior of the  chlorine counterpart of the nabokoite family was reported in \cite{rebrov}, low-frequency (MHz range and below) dielectric properties of K/Cl nabokoite were studied in \cite{markina,china}. We extend here application of these experimental technique to the \nabok{A}{Br}{}  (A=K, Cs, Rb) compounds (Fig.~\ref{fig:dielectric}).

Light-alkali-ion K/Br nabokoite demonstrates strong maximum of low-frequency dielectric permittivity $\varepsilon$ at $(60\pm2)$~K (Fig.~\ref{fig:dielectric}-a). Microwave frequencies experiments revealed dielectric losses that strongly damped the cavity  Q-factor at the same temperature.  No high-temperature dielectric anomalies of comparable strength were observed in  heavy-alkali-ion Cs/Br and Rb/Br compounds.

Weak dielectric absorption at microwaves frequencies was observed in Rb/Br and Cs/Br nabokoites  close to the magnetic transition (see Fig.~\ref{fig:dielectric}-b). Dielectric origin of this absorption is clearly evidenced by comparison of field- and temperature scans at different cavity modes. At TE102 cavity mode sample is in the maximum of microwave magnetic field, usual magnetic resonance absorption signal is observed in a field scan (the observed narrow absorption signal is due to the small amount $\simeq 1\%$ of free paramagnetic centers, see \cite{nabok-pm}) and no absorption around the N\'{e}el point appears in a temperature scan. At TE101 and TE 103 cavity modes sample is in the maximum of microwave electric field, no resonance absorption appears in a field scan, but clear absorption signals are visible in temperature scans. Search for the similar  low-temperature dielectric losses in  K/Br sample is handicapped by the wing of the main dielectric anomaly causing remarkable background de-tuning of the microwave cavity and not allowing to acquire decisive data in our experiments.

Thus, our experimental research demonstrates that high-temperature dielectric anomaly is a characteristic feature of  all  light-alkali-ion nabokoite.

\subsection{Collinear antiferromagnetic order in heavy-alkali-ion nabokoites}

\begin{figure}[th]
\centering
 \includegraphics[width=\columnwidth]{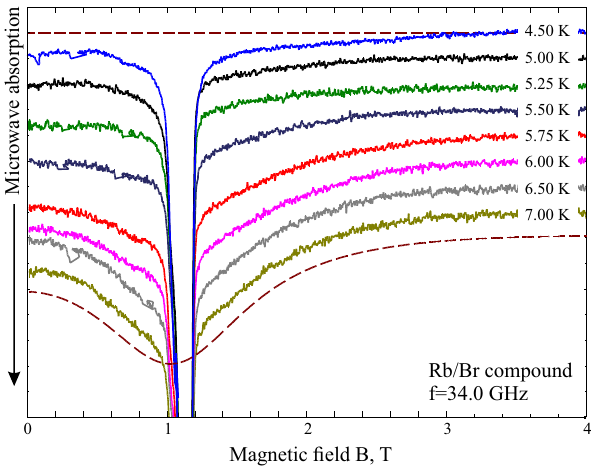}\\
 \caption{(color online) Temperature evolution of the electron spin resonance absorption in the powder sample of  \nabok{Rb}{Br}{} at $f=34.0$~GHz. Dashed curve close to the 7.0~K absorption spectrum is a Lorentzian fit of the ESR absorption line above the N\'{e}el point, curve is shifted for clarity. Horizontal dashed line over the 4.5~K absorption shows ``no absorption'' level to highlight broad absorption band at low fields appearing below the N\'{e}el point. Narrow absorption signal at 1.0...1.2~T is a paramagnetic resonance response from a small amount (about 1\% per copper ion of nabokoite) of free paramagnetic centers \cite{nabok-pm}.}\label{fig:RbBr-spectra(T)}
\end{figure}

\begin{figure}[th]
\centering
 \includegraphics[width=\columnwidth]{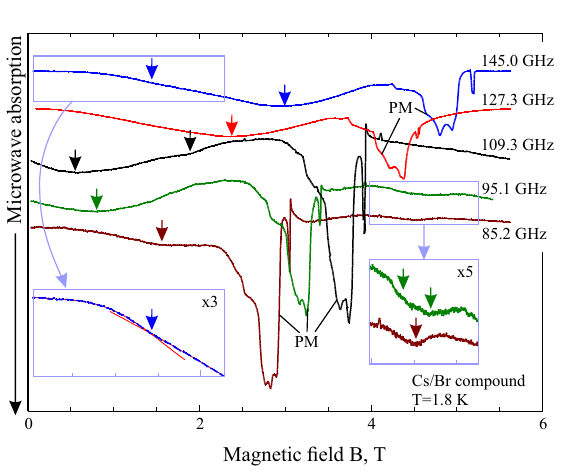}\\
 \caption{(color online) Examples of antiferromagnetic resonance absorption spectra in powder sample of \nabok{Cs}{Br}{} at high microwave frequencies $f>80$~GHz, $T=1.7$~K. Vertical arrows marks positions of the observed absorption features: edges of absorption bands or absorption maxima. Irregularly shaped absorption labeled 'PM' corresponds to the resonance of free paramagnetic centers (about 1\% per copper ion of nabokoite). Narrow absorption to the right from 'PM' absorption is a DPPH marker ($g=2.00$). Insets show magnified fragments of absorption spectra to highlight weak features. Left inset highlights weak kink marking right edge of absorption, red curves are empirical guides-to-the-eye obtained by polynomial fits above and below the kink. Right inset highlights weak absorption in the vicinity the of the spin-flop transition. }\label{fig:100ghz-afmr}
\end{figure}

\begin{figure*}[th]
\centering
  \includegraphics[width=\textwidth]{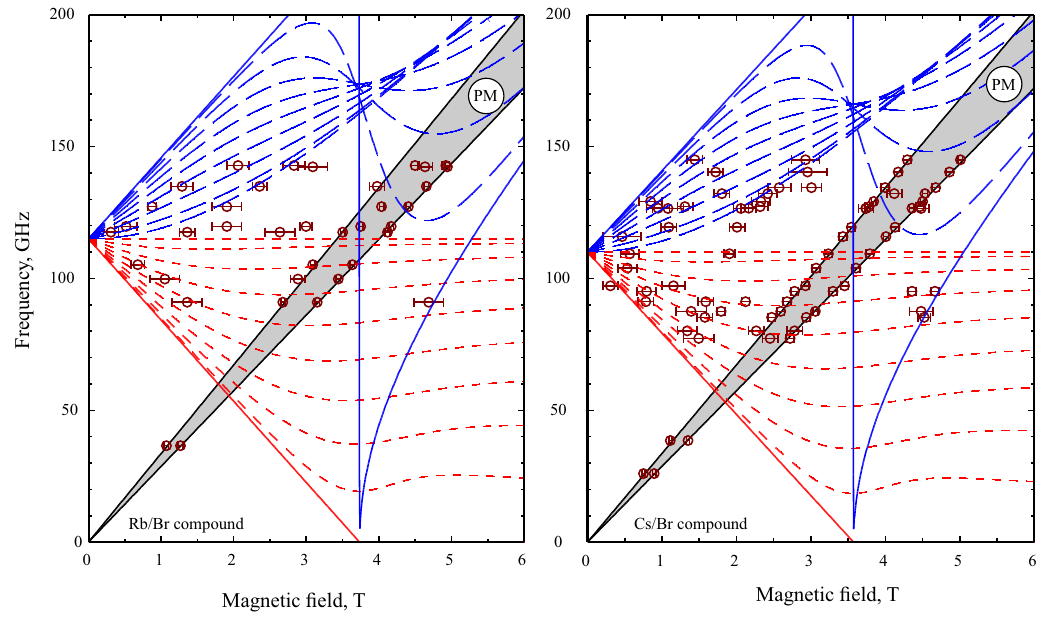}\\
  \caption{(color online) Frequency-field diagrams for antiferromagnetic resonance in powder samples of \nabok{Rb}{Br}{} and \nabok{Cs}{Br}{}, $T=1.7$~K. Symbols --- experimental data taken as characteristic edges of AFMR absorption band, maxima of AFMR absorption and edges of absorption by free paramagnetic centers. Curves --- model $f(B)$ curves for the collinear easy-axis antiferromagnet with the gap $\Delta=115$~GHz for Rb/Br compound and 110~GHz for Cs/Br compound and effective $g$-factor value $g=2.20$. Solid model curves corresponds to the powder particle with anisotropy axis parallel to the field, dashed model curves are calculated for other field directions with 10$^\circ$ step. Greyed area (labeled 'PM') corresponds to the absorption of the free paramagnetic centers, its boundaries correspond to $g$-factor values 2.05 and 2.40. }\label{fig:heavy-afmr}
\end{figure*}

\begin{figure}[th]
\centering
  \includegraphics[width=\columnwidth]{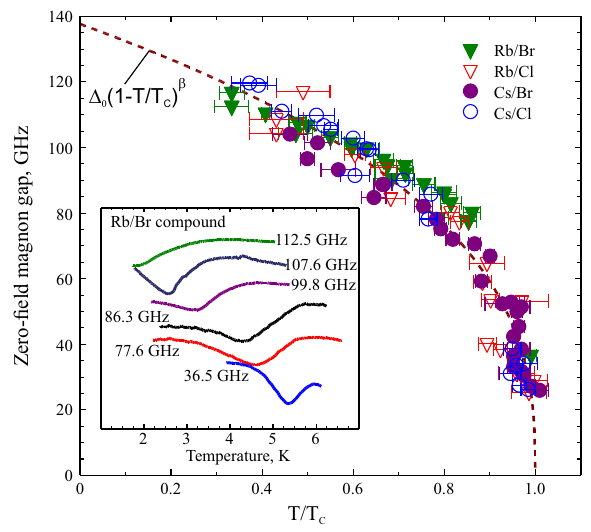}\\
  \caption{(color online) Scaled temperature dependences of zero-field magnon gaps for the heavy-alkali-ion nabokoites. Critical temperature values used for scaling $T_\textrm{C}$ are: 5.40~K for Rb/Br nabokoite, 5.10~K for Rb/Cl, 5.25~K for Cs/Br and 5.10~K for the Cs/Cl compound. Symbols -- experimental data, dashed curve -- empirical critical exponent fit $\Delta(T)=\Delta_0 (1-T/T_\textrm{C})^\beta$ with $\Delta_0=(138\pm3)$~GHz and $\beta=0.37\pm0.05$. Inset: representative examples of the ``temperature resonance'' scans for Rb/Br nabokoite. }\label{fig:tempres-heavy}
\end{figure}

\begin{figure}[th]
\centering
  \includegraphics[width=\columnwidth]{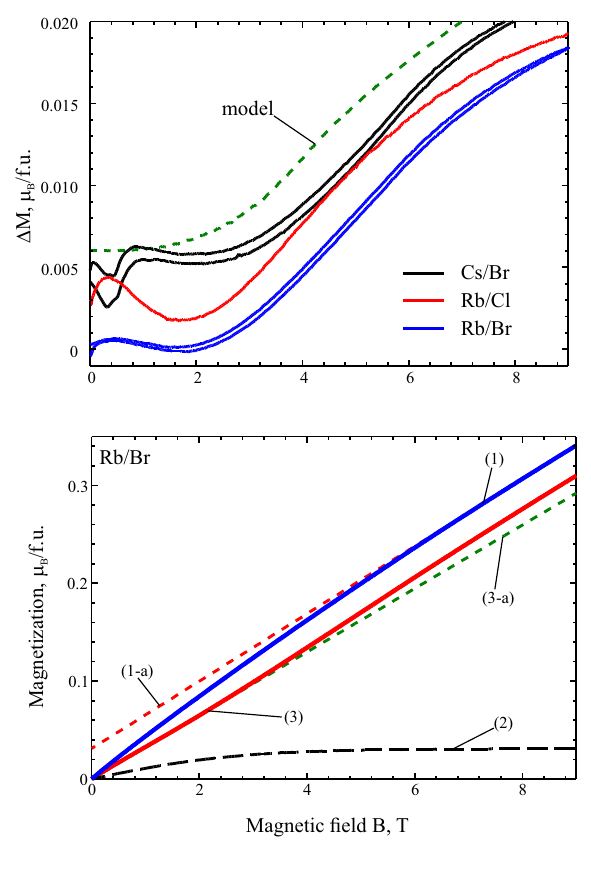}\\
  \caption{(color online) Upper panel:   variation of the antiferromagnetic contribution to sample magnetization $\Delta M(B)$ (see Eqn.~\eqref{eqn:DeltaM})  around the spin-flop field (solid curves) and modeled curve for the powdered easy-axis antiferromagnet with $B_\textrm{SF}=4.0$~T (dotted curve). Lower panel: illustration of the $M(B)$ curve transformation  routine for \nabok{Rb}{Br}. Curves are labeled as follows: (1) original $M(B)$ curve measured at 2~K, (1-a) high-field linear fit of $M(B)$ to provide intercept estimate $m_0$, (2) saturation curve for the ideal paramagnet $M_\textrm{PM}(B)$  with $g=2.20$ and saturation magnetization $m_0$, (3) sample magnetization curve with paramagnetic contribution subtracted $\widetilde{M}(B)=M(B)-M_\textrm{PM}(B)$, (3-a) linear fit of low-field part of $\widetilde{M}(B)$ curve.}\label{fig:heavy-m}
\end{figure}

Temperature evolution of the electron spin resonance absorption at relatively low frequency $f=34.0$~GHz in Rb/Br compound is shown at Fig.~\ref{fig:RbBr-spectra(T)}.

As the Curie-Weiss temperatures in nabokoites are quite large ($\simeq 150$~K), the low temperature ($T<10$~K) paramagnetic susceptibility of the nabokoite matrix spins is strongly reduced compared to the ideal paramagnet. Hence, even a small amount of free paramagnetic centers contribute remarkably both to static and dynamic magnetic properties. Absorption signal observed at 7~K in Rb/Br nabokoite (Fig.~\ref{fig:RbBr-spectra(T)})  consists of two components: a broad paramagnetic line with the linewidth (half-width at half-maximum) $\Delta B=0.68$~T and $g$-factor $g=2.34\pm0.16$ and a narrow absorption line in the field range corresponding approximately  to the $g$-factor values from 2.05 to 2.30.

The narrow absorption signal is due to free paramagnetic centers, its intensity corresponds to approximately 1\% of paramagnetic centers per copper ion \cite{nabok-pm}. Most likely this signal originates from some paramagnetic impurities or powder particles surface defects. Dominating amplitude of the paramagnetic centers signal  (Figs.~\ref{fig:RbBr-spectra(T)},~\ref{fig:100ghz-afmr}) is due to the reduced magnetic susceptibility in the low-temperature paramagnetic state of nabokoites. This signal follows Curie law and is insensitive to the magnetic phase transition.

Broad absorption component, on the contrary, changes at the N\'{e}el point. On cooling below 6~K we observe that spectral density of absorption at $f=34.0$~GHz shifts to lower fields and disappears. Such a behavior is typical to the magnets with the gapped spin-wave spectrum: zero-field magnon gap is proportional to the order parameter value and increases from zero below the transition temperature, as the magnon gap $\Delta (T)$ becomes larger than the microwave energy quantum $2\pi \hbar f$ absorption disappears.

To verify this scenario we extended our experiments to higher frequencies and, indeed, have found low-temperature resonance absorption signals at 80--150~GHz frequency range (Fig.~\ref{fig:100ghz-afmr}). Observed low temperature absorption includes  absorption signal from paramagnetic centers labeled 'PM', broad absorption band to the left from the paramagnetic absorption and a weak additional absorption signal around 4.0~T. All four heavy-alkali-ion nabokoites (Rb/Cl, Rb/Br, Cs/Cl and Cs/Br compounds) demonstrate similar behavior.

Broad absorption band to the left of the paramagnetic absorption is regularly observed from 80 to 150~GHz. Antiferromagnetic resonance (AFMR) absorption is known to be anisotropic \cite{kubo,gurevich}, the observed absorption band corresponds to the powder averaging of AFMR absorption with left and right boundaries corresponding to some limiting powder particles orientations and  a maximum of absorption determined by the  statistical weight in distribution of AFMR eigenfrequencies and anisotropic absorption intensity. The absorption boundaries are smoothed by the finite linewidth of the AFMR absorption turning sharp edges of absorption  into kink-like features. We took absorption maxima and kinks at the edges of absorption bands as characteristic fields of AFMR absorption (see arrows and insets at Fig.~\ref{fig:100ghz-afmr}) and plot collected data at frequency-field diagrams (Fig.~\ref{fig:heavy-afmr}). Boundaries of absorption by free paramagnetic centers are also plotted at  Fig.~\ref{fig:heavy-afmr} to shade out area of the frequency-field diagram where paramagnetic centers absorption blocks AFMR signal observation.

Observed distributions of the AFMR spectra features are in a good agreement with the powder-averaged AFMR absorption in a collinear antiferromagnet with easy-axis anisotropy with the zero-field magnon gap $\Delta=115$~GHz for Rb/Br compound and 110~GHz for Cs/Br compound (both values correspond to temperature $T=1.7$~K).

Computed $f(B)$ dependences for different angles between the magnetic field and anisotropy axis \cite{kubo, gurevich} are shown at Fig.~\ref{fig:heavy-afmr} with 10$^\circ$ step assuming mean $g$-factor value $g=2.20$ and the above gap values.  Agreeing with the experimental observations, the  computed curves clearly show that at frequencies $f<\Delta$ only the left edge of absorption band is expected while the right edge is undefined while at $f>\Delta$ the left edge of absorption band  corresponds to rare powder particles with the anisotropy axis parallel to the field and statistically more probable powder particles with anisotropy axis orthogonal to the field provide well defined right edge of absorption band.

Observed magnon gaps correspond to the spin-flop fields $B_c=\Delta/\gamma$ \cite{kubo, gurevich} of 3.7 and 3.6~T for Rb/Br and Cs/Br compounds  correspondingly (with approximately 10-15\% uncertainties arising both from the gap estimation uncertainty and uncertainty of the  $g$-factor value for the field applied along the easy axis). Note, that the weak absorption signals observed to the right from the paramagnetic resonance position fits well to the model predictions, weakness of these signals and difficulty of their regular observation is due to small statistical weight of the corresponding powder particles. Appendix~\ref{sec:app-pow} includes more detailed modeling of powder particles AFMR eigenfrequencies distribution for the case of easy-axis, easy-plane and bi-axial anisotropy which proves that only the easy-axis anisotropy fits to the observed $f(B)$ patterns.

While analysis of the AFMR absorption is complicated by powder averaging of the anisotropic resonance signal, zero-field ``temperature resonance'' experiment turns out to be very informative and reliable in interpretation as the powder nature of the sample is of no importance at zero field.

We have found that temperature scans of microwave absorption reveal regular temperature dependent-absorption signal (Fig.~\ref{fig:tempres-heavy}). This absorption appears once the spin-wave gap equals the microwave frequency and ``one photon absorption -- one magnon excitation'' processes became allowed. Thus, temperature dependent zero-field gaps $\Delta(T)$ can be followed with high accuracy from low temperatures up to the transition temperature $T_\textrm{C}$. Observed temperature dependence of the gap can be empirically fitted over the entire temperature range with equation

\begin{equation}
\label{eqn:ea-gap(T)}
\Delta(T)=\Delta_0 \left(1-T/T_\textrm{C}\right)^\beta
\end{equation}

\noindent Transition temperatures $T_\textrm{C}$ determined from independent fits of $\Delta(T)$ data are 5.40~K for Rb/Br nabokoite, 5.10~K for Rb/Cl, 5.25~K for Cs/Br and 5.10~K for the Cs/Cl compound (Table~\ref{tab:temperatures}). Within our experiment accuracy, the scaled zero-field gap temperature dependences $\Delta(T/T_\textrm{C})$ follow the same curve \eqref{eqn:ea-gap(T)}  for all four heavy-alkali-ion nabokoites (Fig.~\ref{fig:tempres-heavy}). The empirical critical exponent value $\beta=0.36\pm0.05$ is close to the values expected for the 3D antiferromagnets \cite{exp1,exp2,exp3} and is well away from the 2D Ising model critical exponent 1/8. This observation supports formation of the conventional three-dimensional antiferromagnetic ordering below the N\'{e}el temperature.

Spin-flop transition in a single-crystalline sample is usually marked by sharp increase of magnetization \cite{kubo}. In our case spin-flop signature on a magnetization curve is smeared by powder averaging, non-linear contribution of free paramagnetic centers, contribution from  not ordered spin subsystem \cite{nabok-pm}.
However, these contributions can be reasonably estimated since magnetization of free paramagnetic centers at 2~K saturates at $B=1...2$~T as $$M_\textrm{PM}(B)=m_0 \tanh\left(\frac{g \mu_\textrm{B} B}{2 k_\textrm{B} T}\right)$$ and possible contribution of the magnetically decoupled not ordered spin subsystem is expected  to be linear at $B<10$~T because of the large exchange coupling parameters.

We use following routine. Firstly, we fit the high-field (8-9~T, i.e. well above both the spin-flop field and the saturation field ) part of $M(B)$ curve $$M(B)=m_0+\alpha B$$ Intercept $m_0$ can be interpreted as the free paramagnetic centers saturated magnetization. Secondly, we compute saturation curve for the free paramagnetic centers  $M_\textrm{PM}(B)$ with the found value of $m_0$ and  mean $g$-factor value $g=2.20$. Finally, we fit the  low-field (1-2~T, i.e. reasonably below the spin-flop field) part of residual magnetization curve $$\widetilde{M}(B)=M(B)-M_\textrm{PM}(B)$$ by a linear law $$\widetilde{M}(B)=\xi B$$ This linear law includes magnetization contributions  from finite longitudinal susceptibility of easy axis antiferromagnet at finite temperature, powder averaging effects and possible contribution from the not ordered magnetic subsystems.

The difference
\begin{equation}
\label{eqn:DeltaM}
\Delta M(B)=\widetilde{M}(B)-\xi B
\end{equation}

\noindent can be interpreted as a change of magnetization caused by spin-reorientation transition.

This routine was successfully repeated for Cs/Br, Rb/Cl and Rb/Br compounds (Fig.~\ref{fig:heavy-m}).  Curves $\Delta M(B)$ for all three compounds coincides within error margins with each other and with the modeled curve for the powdered easy-axis antiferromagnet with $B_\textrm{SF}=4.0$~T. Value of magnetization change $\Delta M$ at a spin-flop field $\Delta M \approx 0.005 \mu_\textrm{B}/\textrm{f.u.}$ allows to estimate antiferromagnetic phase susceptibility (see Eqn.~\eqref{eqn:dM(Hsf)})

\begin{equation}
\label{eqn:chi-estim}
\chi_\perp\approx\frac{\Delta M(H_\textrm{SF})}{0.22 H_\textrm{SF}} \approx 0.0032~\textrm{EMU/mole}
\end{equation}

\noindent this value can be compared with the conventional estimate of collinear antiferromagnet transverse susceptibility $\mu_\textrm{B}^2 N_\textrm{A}/J$ yielding estimate of mean exchange coupling of the ordered subsystem $J/k_\textrm{B} \simeq 100$~K which is reasonably close to the estimated couplings in nabokoites \cite{DFT}.

Thus, we have found that the magnetic ordering in heavy-alkali-ion nabokoites below the N\'{e}el point is the conventional 3D collinear easy-axis antiferromagnetic order. Characteristics of the ordered phase (zero-field magnon gaps and spin-flop fields) are found to be practically insensitive to the details of nabokoite chemical composition.

\subsection{Two-step magnetic phase transition into the noncollinear antiferromagnetic state in light-alkali-ion nabokoites}
\begin{table*}[th]
\caption{Characteristics of the noncollinear antiferromagnetic order in light-alkali-ion nabokoites \nabok{A}{X}{} (A=Na, K) obtained from electron spin resonance experiments: magnon gaps $\Delta_{1,2,3}$ at 1.7~K (10~GHz corresponds to 0.48~K or 0.041~meV), critical fields $B_\textrm{c1,c2,c3}$, anomalous slope factor $\Gamma$ (see text for definition).}
\label{tab:noncoldata}
\begin{ruledtabular}
\begin{tabular}{cccccccc}
Compound& $\Delta_{1}$, GHz& $\Delta_{2}$, GHz& $\Delta_{3}$, GHz&$B_\textrm{c1}$, T&$B_\textrm{c2}$, T&$B_\textrm{c3}$, T&$\Gamma$\\
\hline
Na/Cl&$27.5\pm 2.0$&$112\pm 5$&$153\pm 5$&$0.55\pm0.10$&$1.9\pm 0.2$&$4.0\pm0.5$&$1.74\pm0.09$\\
K/Cl\cite{markina}&$18.0\pm 1.0$&$75\pm 3$& no data &$0.230\pm0.015$&$0.80\pm0.10$& no data&$2.57\pm0.09$\\
K/Br&$13.0\pm 3.0$&$73\pm 3$& no data&$0.22\pm 0.05$&$0.83\pm 0.05$& no data&$2.61\pm0.09$\\
\end{tabular}
\end{ruledtabular}
\end{table*}

\begin{figure}[th]
\centering
  \includegraphics[width=\columnwidth]{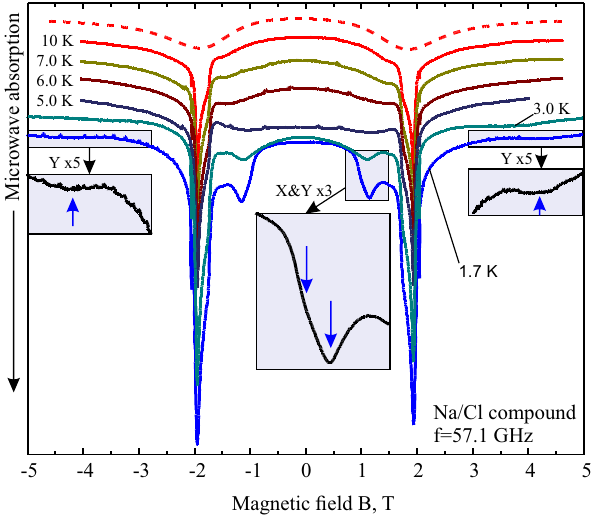}\\
  \caption{(color online) Temperature evolution of the electron spin resonance absorption in the powder sample of  \nabok{Na}{Cl}, $f=57.1$~GHz. Solid curves -- experimental data (note the bi-polar field sweeps). Dashed curve -- Lorentzian fit of 10~K absorption spectra with excluded paramagnetic centers contribution. Fragments of 1.7~K absorption curves are magnified to highlight characteristic features of absorption spectra: weak absorption at approx.~4~T (Y-magnified five-fold) and absorption band at low fields corresponding to the AFMR mode with anomalous slope (X- and Y-magnified three-fold). Vertical arrows mark positions of the characteristic absorption features. }\label{fig:NaCl-spectra(T)}
\end{figure}

\begin{figure*}[th]
\centering
  \includegraphics[width=\textwidth]{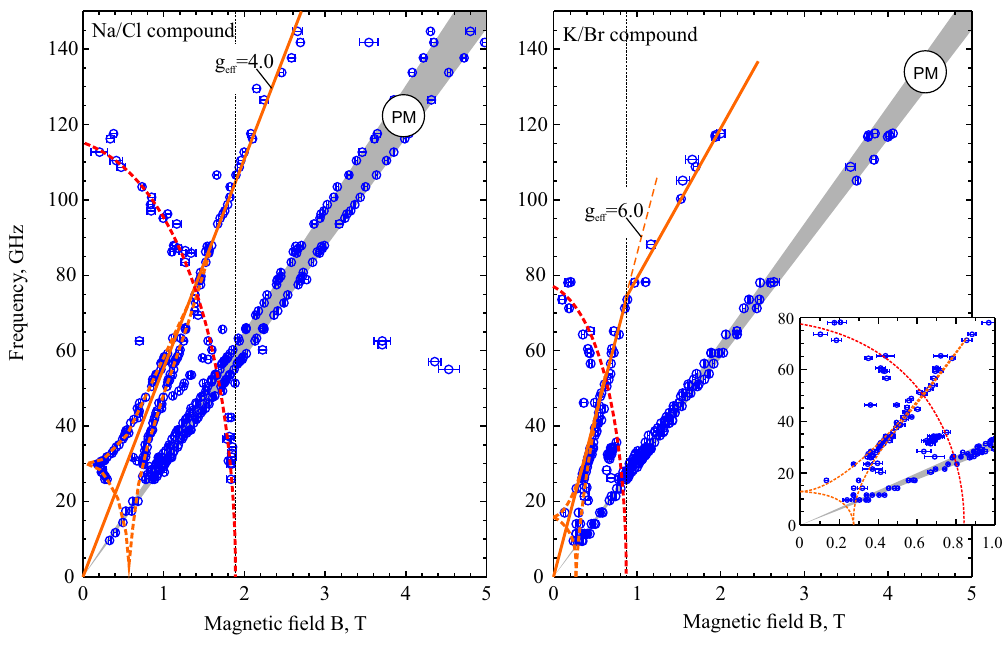}\\
  \caption{(color online) Frequency-field diagrams for \nabok{Na}{Cl}{} (left) and \nabok{K}{Br} (right) at 1.7~K. The inset at the right panel show expanded low-field part of the frequency-field diagram for K/Br compound. Symbols -- experimental data (characteristic edges of absorption bands and absorption maximums).  Greyed area (labeled 'PM') corresponds to the absorption of the free paramagnetic centers, its boundaries correspond to $g$-factor values 2.10 and 2.35 for Na/Cl compound and $g$-factor values 2.07 and 2.25 for K/Br compound.  Solid (orange) lines show linear asymptotes for non-Larmor AFMR modes. Dashed (red and orange) curves  are guides to the eye showing softening of the gapped AFMR mode at critical fields $B_\textrm{c1}$ and $B_\textrm{c2}$. Vertical (black) dashed lines mark second critical field $B_\textrm{c2}$.}\label{fig:light-afmr}
\end{figure*}

\begin{figure}[th]
\centering
  \includegraphics[width=\columnwidth]{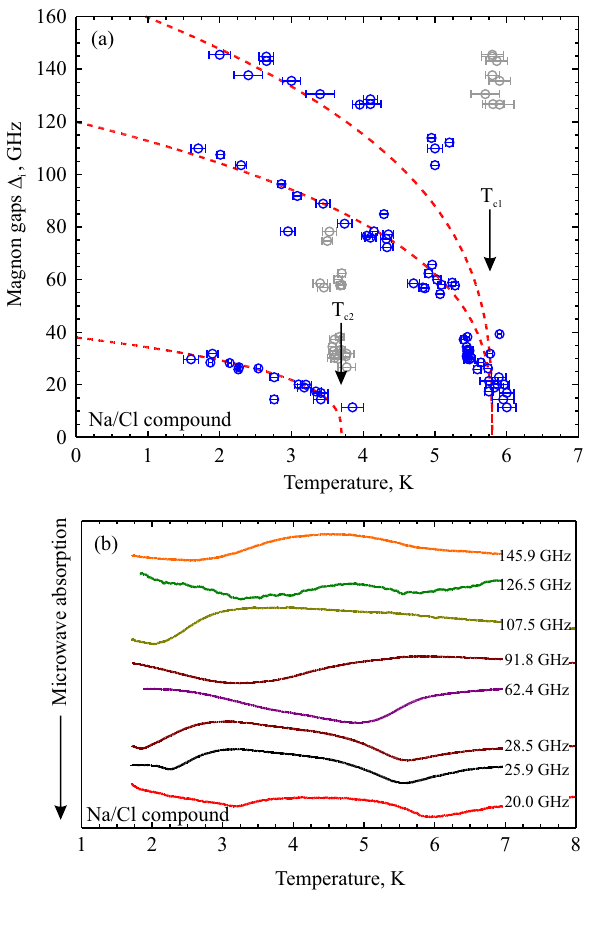}\\
  \caption{(color online) (a) Temperature dependence of the zero-field magnon gaps in \nabok{Na}{Cl}. Symbols  -- experimental data, dashed curves are guides to the eye. Blue symbols mark magnon datapoints, grey symbols mark kinks or weak features at the transition temperatures. (b) Representative examples of the ``temperature resonance'' scans.}\label{fig:light-gap}
\end{figure}

\begin{figure}[th]
\centering
  \includegraphics[width=\columnwidth]{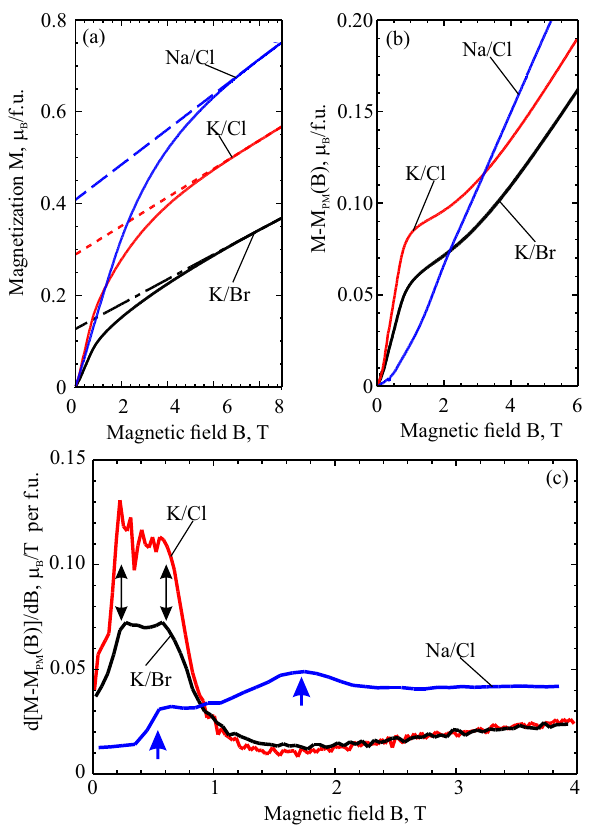}\\
  \caption{(color online) (a) Solid curves -- low-temperature (2~K) magnetization curves for the K/Cl, K/Br and Na/Cl-nabokoites. Dashed lines -- linear fits of the high-field (above 7~T) part of magnetization curves to determine saturated magnetization of the free paramagnetic centers. (b) Residual magnetization  $\widetilde{M}=M-M_\textrm{PM}(B)$, contribution of free paramagnetic centers calculated for $g=2.20$ and saturation magnetization $m_0$ determined from panel (a), $T=2$~K. (c) Field derivatives of the residual magnetization $d\widetilde{M}/dB$, vertical arrows mark determined critical fields $B_\textrm{c1}$ and $B_\textrm{c2}$.}\label{fig:m-light}
\end{figure}

Antiferromagnetic resonance absorption in light-alkali-ion nabokoites (K/Cl \cite{markina}, K/Br and Na/Cl compounds)  is qualitatively different from that in heavy-alkali-ion nabokoites. Representative temperature evolution of the resonance absorption for Na/Cl compound is shown at Fig.~\ref{fig:NaCl-spectra(T)}. A broad paramagnetic resonance absorption (half-width at half-maximum $\Delta B=0.72$~T at 10~K) transforms at the N\'{e}el point into a complicated absorption spectrum with several absorption bands. Well defined antiferromagnetic resonance absorption is observed at frequencies below 100~GHz and the resulting frequency-field diagrams look very different from those discussed above (compare Fig.~\ref{fig:heavy-afmr} and Fig.~\ref{fig:light-afmr}): there are two zero-field magnon gaps $\Delta_1$ and $\Delta_2$ clearly visible at $f(B)$ diagram, the magnetic resonance modes originating from these modes soften at different critical fields $B_\textrm{c1}$ and $B_\textrm{c2}$. Additionally, weak absorption at the field around 4~T was observed in Na/Cl compound (Fig.~\ref{fig:NaCl-spectra(T)}) probably indicative of the third critical field $B_\textrm{c3}$. Values of magnon gaps and critical fields are collected in Table~\ref{tab:noncoldata}.

The most characteristic feature of the $f(B)$ diagrams for all light-alkali-ion nabokoites is the non-Larmor asymptotic behavior of one of the AFMR modes. The slope of this mode corresponds to the effective $g$-factor values  $g_\textrm{eff}=4.0\pm0.2$ for Na/Cl nabokoite and $g_\textrm{eff}=6.0\pm0.2$ for K/Br-nabokoite, a close value $g_\textrm{eff}=5.9\pm0.2$ was reported earlier \cite{markina} for K/Cl-compound. These $g$-values are very unusual for Cu$^{2+}$ ions, paramagnetic resonance absorption in all nabokoites above the N\'{e}el temperature can be described by mean $g$-factor value  $g_\textrm{PM}=2.3\pm0.1$. We will characterize slope of this non-Larmor mode by an anomalous slope factor

\begin{equation}
\label{eqn:gamma-factor}
\Gamma=\frac{g_\textrm{eff}}{g_\textrm{PM}}
\end{equation}

\noindent which is about 1.8 for Na/Cl-nabokoite and about 2.6 for K/Cl and K/Br-compounds (see Table~\ref{tab:noncoldata}). Such non-Larmor mode is a signature of noncollinear antiferromagnetic ordering. Note also, that a change of slope of this non-Larmor mode  is observed in K/Br-nabokoite at $B_\textrm{c2}$ (Fig.~\ref{fig:light-afmr}).

``Temperature resonance'' experiment was successful for Na/Cl compound, which has the highest ferroelectric transition temperature (Table~\ref{tab:temperatures}), see Fig.~\ref{fig:light-gap}.  Some of the absorption features in ``temperature resonance'' experiment turn out to be temperature independent (see grey data points at Fig.~\ref{fig:light-gap}) which is most likely the effect of the phase transition itself yielding some changes in microwave magnetic or dielectric properties of the sample. However, temperature dependence of  three magnon gaps $\Delta_1<\Delta_2<\Delta_3$ can be followed reliably. The largest gaps $\Delta_2$ and $\Delta_3$ open at $T_\textrm{c1}=5.8\pm0.2$~K, while the smallest gap $\Delta_1$ appears at the lower temperature $T_\textrm{c2}=3.7\pm0.1$~K. This observation directly proves that magnetic ordering in Na/Cl-nabokoite is established via two-step phase transition and confirms magnetic origin of the specific heat peaks (Fig.~\ref{fig:specheat}). Temperature dependence of all gaps can be fitted by  empirical law

\begin{equation}\label{eqn:gaps-noncol}
\Delta_i(T)=\Delta_{i0}\left(1-T/T_\textrm{C}\right)^\beta
 \end{equation}
 \noindent with the same critical exponent $\beta=0.33$ (the value, again, typical for 3D magnetic ordering \cite{exp1,exp2,exp3}) and gaps $\Delta_{10}=38$~GHz, $\Delta_{20}=120$~GHz and $\Delta_{30}=170$~GHz.

 Dielectric anomalies in K-containing nabokoites are located at lower temperatures, which  results in the strong temperature-dependent microwave cavity de-tuning at the temperatures of interest preventing accurate determination of magnon gaps be means of ``temperature resonance'' technique.

Thus, both specific heat data (Fig.~\ref{fig:specheat}) and ``temperature resonance'' experiments (Fig.~\ref{fig:light-gap}) demonstrate presence of two successive phase transitions in light-alkali-ion nabokoites. It is important to note (see Fig.~\ref{fig:NaCl-spectra(T)}) that absorption corresponding to the non-Larmor mode (AFMR absorption at the fields below paramagnetic resonance field) develops below the upper transition temperature $T_\textrm{c1}$ (see, e.g., 5~K absorption spectra at Fig.~\ref{fig:NaCl-spectra(T)}). We have checked in the separate experiment at $T=5.0$~K that non-Larmor mode is regularly observed at $T_\textrm{c2}<T<T_\textrm{c1}$ and its anomalous slope factor $\Gamma(\textrm{5 K})$ is approximately the same as at 1.7~K.

Critical fields  $B_\textrm{c1}$ and $B_\textrm{c2}$ can be also extracted from low-temperature magnetization curves (Fig.~\ref{fig:m-light}) (see also \cite{alisher,markina}). Since magnetization $M(B)$ includes contribution from free paramagnetic centers we, again, fit the high-field part (above 7~T) of the low-temperature magnetization curve linearly and take the intercept $m_0$ as estimate of saturated magnetization of free paramagnetic centers (Fig.~\ref{fig:m-light}-b). Afterwards we calculate residual magnetization $\widetilde{M}=M-M_\textrm{PM}(B)$  (Fig.~\ref{fig:m-light}-b). The critical fields can be visualized by taking the field derivative of the residual magnetization $d\widetilde{M}/dB$, see Fig.~\ref{fig:m-light}-c which demonstrates two peaks. We take derivative maximums as marks of the critical fields, for  both K/Cl and K/Br-nabokoites $B_\textrm{c1}=(0.23\pm0.05)$~T and  $B_\textrm{c2}=(0.60\pm0.15)$~T, while for the Na/Cl compound $B_\textrm{c1}=(0.55\pm0.10)$~T and $B_\textrm{c2}=(1.70\pm0.15)$~T. All values are close to the values determined form $f(B)$ diagrams (see Table~\ref{tab:noncoldata} and Fig.~\ref{fig:light-afmr}).

Summing up our experimental findings for the light-alkali-ion nabokoite family compounds, we observe that magnetically ordered phase in light-alkali-ion nabokoites \nabok{A}{X}{}, A=Na,~K is a noncollinear antiferromagnetic ordering formed via two-step phase transition. Characteristics of the ordered phase (magnon gaps, critical fields) are individual for all light-alkali-ion nabokoites.

\section{Discussion}
\begin{figure}[th]
\centering
  \includegraphics[width=\columnwidth]{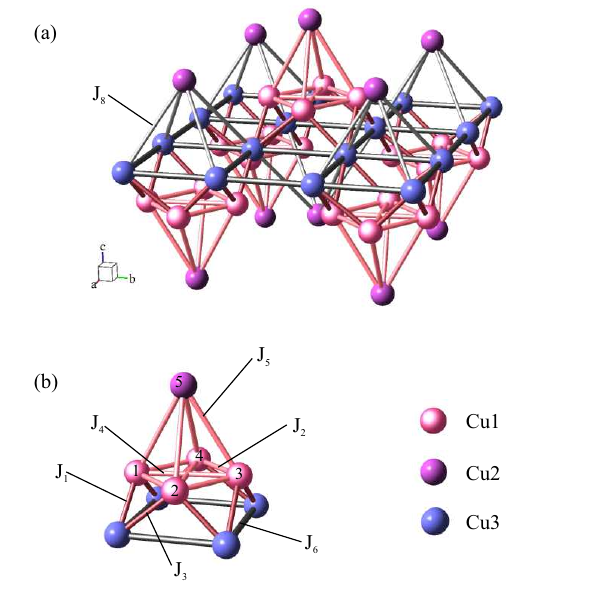}\\
  \caption{(color online) Relevant exchange bonds in the decorates SKL lattice of K/Cl nabokoite  according to DFT calculations \cite{DFT}. Numeration of exchange bonds follows \cite{DFT}. Only positions of copper ions are shown for clarity. (a) 2D layer with decorating copper ions.. (b) Single pyramidal element, spins at the pyramid vertices are enumerated as in Eqn.~\eqref{eqn:pyramid-ham}.}\label{fig:j-bonds}
\end{figure}

\subsection{Ordered phases of nabokoites}

Characteristic transformation of the electron spin resonance absorption spectra at the N\'{e}el temperature confirms formation of the low-temperature antiferromagnetic order in nabokoite family compounds. ``Temperature resonance'' experiments (Figs.~\ref{fig:tempres-heavy},~\ref{fig:light-gap}) demonstrate that magnon gaps, which are proportional to the antiferromagnetic order parameter, develops continuously from zero, as expected for the second order phase transitions. Empirical critical exponents \eqref{eqn:ea-gap(T)}~,\eqref{eqn:gaps-noncol}, determined over broad temperature range, are close to the values expected for 3D magnetic ordering. Thus, our experimental observations are in agreement with the formation of 3D magnetic ordering in nabokoite family compounds at low temperatures.

Spin-waves in a Heisenberg antiferromagnet have linear dispersion law $\varepsilon(k)\propto k$ which yields for specific heat, analogously to Debye theory, $C_\textrm{3D}(T)\propto T^3$ for 3D antiferromagnet. Experimental data (Fig.~\ref{fig:specheat}) demonstrates that even at $T<T_\textrm{N}/2$ temperature dependence of the specific heat is more close to 2D law $C_\textrm{2D}(T)\propto T^2$. Such a behavior is known for quasy-2D antiferromagnets \cite{svistov-cross}: for the weak coupling between 2D subsystems spread of the dispersion $\varepsilon(\vec{k})$ is very anisotropic and 3D to 2D crossover in specific heat temperature dependence occurs  at temperature close to the inter-layer coupling strength  $T*=J_\perp/k_\textrm{B}$. Therefore, observed temperature dependence of the specific heat points to quasy-2D character of the ordering spin system in nabokoites.

Comparison of spin susceptibilities determined from the static measurements and from the magnetic resonance \cite{nabok-pm} provides indications that only fraction of spins in nabokoites (1--3 copper ions per nabokoite formula unit) form magnetically ordered state. Our data alone cannot determine either microscopic location of the ordered spin subsystem nor the microscopic pattern of sublattices mutual orientation, such identification requires microscopic probes (e.g.,  NMR or $\mu$SR) or magnetic neutron diffraction. This microscopic problem is additionally complicated by nabokoite crystal structure (Fig.~\ref{fig:struct}) providing an entangled network of possible spin-spin bonds with seven magnetic ions per formula unit and four formula units per high-temperature unit cell.

However, structural considerations provides two plausible models of partial magnetic ordering in nabokoites with quasy-2D character of the ordered spin subsystem:

(i) Assuming nearest neighbor interactions only  (Fig.~\ref{fig:struct}), nabokoites spins can be naturally divided into 2D square kagom\'{e} layers and decorating interlayer magnetic ions. This leads to the model proposed in \cite{markina} combining subsystems of disordered (spin-liquid) SKL layers and decorating interlayer ions. The later can order due to effective coupling mediated by virtual excitations of 2D spin-liquid layers.

(ii) Recent DFT calculations for K/Cl-nabokoite \cite{DFT} suggest importance of the next-nearest neighbors interactions. The suggested bonds structure is shown at Fig.~\ref{fig:j-bonds}, the dominating exchange coupling is expected along diagonals of Cu1 ions squares ($J_4/k_\textrm{B}=170$~K, bonds numeration follows \cite{DFT}), largest of Cu1-Cu3 couplings is $J_3=0.971J_4$, coupling along the side of Cu1-Cu1 square $J_2=0.694 J_4$ and there is a strong coupling Cu1-Cu2 to the interlayer (decorating) ion  $J_5=0.894 J_4$. What is specially important for our discussion, the Cu3-Cu3 next-nearest neighbor bonds forming 2D square lattice are predicted to be non-negligible $J_6=0.182 J_4$ ($J_6/k_\textrm{B}=31$~K). The remaining couplings $J_1$ and $J_8$ are predicted to be much smaller. Formation of the simple square lattice of Cu3 ions and strong couplings within the Cu1-Cu2 pyramid suggests different scenario of subsystems separation: Cu3 ions form simple square lattice ordering antiferromagnetically, while Cu1-Cu2 pyramids provide checkerboard-like decorating motives with disordered or partially ordered spins.

Existing experimental data, including single-crystalline experiments \cite{china}, do not allow to chose any of theses  scenarios definitely. In one of the following subsections we will discuss  that second scenario (square lattice decorated with pyramidal blocks) provides possible clue to the  behavior of the nabokoites low-temperature ordered phases.

\subsection{Antiferromagnetic resonance in the noncollinear ordered phase of light-alkali-ion nabokoites}

The most characteristic feature of magnetic resonance in light-alkali-ion nabokoites is the non-Larmor antiferromagnetic resonance mode (Fig.~\ref{fig:light-afmr}). AFMR in collinear antiferromagnets provides no such mode \cite{kubo, gurevich}, while noncollinear antiferromagnets are known to provide AFMR mode with non-Larmor slope \cite{svistfar,markina,noncolnum} (see also Appendix \ref{sec:app-noncol}).

Low-frequency spin dynamics of the noncollinear antiferromagnet can be described within hydrodynamic approach \cite{andmar,svistfar}. Within this approach spin structure at $T=0$ is considered as being rigidly fixed by exchange couplings, which allows to describe system with arbitrary number of sublattices by no more then three mutually orthogonal unit vectors ${\vec{l}}_{1,2,3}$. The spin vector at given crystallographic site ${\vec{r}}_i$ can be expressed via these antiferromagnetic vectors for the established ordering pattern. E.g., for a helical magnetic structure with propagation vector $\vec{q}$ antiferromagnetic vectors $\vec{l}_1$ and $\vec{l}_2$ determine the spin helix
\begin{equation}\label{eqn:spiral}
\vec{S}({\vec{r}}_i)={\vec{l}}_1 \cos(\vec{q}{\vec {r}}_i)+{\vec{l}}_2 \sin(\vec{q}{\vec{r}}_i)
 \end{equation}
\noindent and the third vector can be defined as normal to the spin helix plane  ${\vec{l}}_3=[{\vec{l}}_1 \times{\vec{l}}_2]$. Uniform ($k=0$) oscillations of spin structure correspond then to the oscillations of ${\vec{l}}_{1,2,3}$ basis. The eigenfrequencies of these oscillations can be found via Euler-Lagrange equation taking into account appropriate form of anisotropy energy and magnetic susceptibility tensor \cite{svistfar,noncolnum}.

The non-Larmor mode characteristic for the noncollinear structures originates from the difference of static susceptibilities for the field applied differently with respect to the ordered spin structure (see Appendix~\ref{sec:app-noncol} for details, Eqn.~\eqref{eqn:omega-no-anis},~\eqref{eqn:app-gamma-chi}) even in Heisenberg case. This means, that asymptotic behavior of this resonance mode should be independent of  powder particle orientation. This suppresses effect of powder averaging and makes this mode well defined and easy to observe. Indeed (Fig.~\ref{fig:light-afmr}), one can see that the non-Larmor mode becomes more narrow as its resonance field increases.

Assuming for the sake of simplicity $\chi_3>\chi_2=\chi_1$ (here $\chi_i$ is the magnetic susceptibility for the field applied along ${\vec{l}}_i$)  observed anomalous slope factor $\Gamma=1.74$ for Na/Cl nabokoite corresponds to $\chi_3/\chi_1=2.74$ and $\Gamma\approx2.6$ for K/Cl and K/Br compounds correspond to $\chi_3/\chi_1\approx3.6$.

Oscillations of ${\vec{l}}_{1,2,3}$ basis can be parameterized by three variables (e.g., Euler angles), which yields three low-energy modes of spin oscillations. We observe three zero-field magnon gaps in light-alkali-ion nabokoites (Fig.~\ref{fig:light-gap}). Description of the similar AFMR frequency-field diagrams for \nabok{K}{Cl}{} was discussed in \cite{markina} (see Supplementary materials therein for details) and this analysis revealed for the case of axial anisotropy susceptibility tensor $\chi_3>\chi_2=\chi_1$ that non-Larmor mode always arises from the largest of anisotropy-induced zero-field magnon gaps. We performed tentative simulations of frequency-fields diagrams for the case of general anisotropic susceptibility tensor $\chi_3>\chi_2>\chi_1$ using numerical approach of Ref.~\cite{noncolnum,numa-scripts} which yields similar results: the non-Larmor mode always arises from the largest gap. Thus, as it was already proposed in \cite{markina}, only the smallest gap $\Delta_1$ should be considered as anisotropy-induced gap, while larger gaps $\Delta_{2,3}$ are of exchange origin. Minimal model providing single anisotropic gap is the spiral magnetic structure with easy-axis anisotropy for the ${\vec{l}}_3$ vector (this vector is normal to the spiral plane), as was discussed in \cite{markina}.

Consequently, the observed critical fields have to be interpreted differently. The lower critical field $B_\textrm{c1}$ is the spin-flop field, corresponding to the rotation of the spiral plane if the applied field is applied within the plane (normally to ${\vec{l}}_3$). Note, that contrary to the spin-flop transition in collinear easy-axis antiferromagnet, spin-flop transition for powdered helimagnet is observed in statistically significant fraction of powder particles, which explains why it is well visible in experiment. The upper critical field $B_\textrm{c2}$ corresponds to the softening of one of the exchange modes with zero-field magnon gap $\Delta_2$, i.e. to the  change of the spin structure. This conclusion is supported by the change of the slope of non-Larmor mode observed in \nabok{K}{Br}{} at second critical field (Fig.~\ref{fig:light-afmr}) since the change of the slope means modification of susceptibility tensor eigenvalues.

Finally, we comment the temperature dependence of the magnon gaps with two-steps phase transition (Fig.~\ref{fig:light-gap}). Similar scenarios were observed in other frustrated magnets: successive phase transitions in pyrochlore antiferromagnet Gd$_2$Ti$_2$O$_7$ are explained in terms of partially disordered high-temperature phase \cite{GTO1}, two-stage transition in a quasy-1D antiferromagnet LiCu$_2$O$_2$ frustrated by next-nearest neighbor exchange coupling was interpreted as successive formation of spin-modulated incommensurate structure in the intermediate phase and a helical magnetic structure in the low-temperature phase  \cite{svistov-licuPRB}. Recent theoretical study demonstrates instability of the frustrated magnets on the tetragonal lattice towards formation of 2-k or hybrid structures with similar two-stage phase transition pattern \cite{utesov}.

We observe that the lowest magnon gap \emph{opens or significantly increases} at the lower transition temperature. As discussed above, high-frequency modes $\Delta_2$ and $\Delta_3$ are of exchange origin, while the low-frequency $\Delta_1$ mode is the former Goldstone mode acquiring a gap due to some anisotropic spin-spin interactions. Since the lattice symmetry of nabokoite is below cubic, gapless behavior of this mode at $T_\textrm{c2}<T<T_\textrm{c1}$ is surprising. We cannot exclude the possibility  that this AFMR mode appears already at the higher transition temperature $T_\textrm{c1}$, but the zero-field magnon gap remains below 10~GHz. The non-Larmor mode appears already at higher transition temperature $T_\textrm{c1}$ (Fig.~\ref{fig:NaCl-spectra(T)}) and its asymptotic slope (anomalous slope factor $\Gamma$)  does not change much at the lower temperature phase transition.

Quasi-low-dimensionality of nabokoite spin subsystems (see Figs.~\ref{fig:struct},~\ref{fig:j-bonds}) allows to hypothesize that lower-temperature transition can be incommensurate-commensurate transition in interlayer ordering pattern. Such a commensurate-incommensurate interlayer ordering pattern transition  was observed, e.g.,  in a quasy-2D triangular lattice magnet RbFe(MoO$_4$)$_2$ \cite{kenzelman}.
Assuming for the sake of simplicity that only nearest neighbors anisotropic spin-spin interactions contributes to the anisotropy energy and that the 2D layers are exactly stacked one-atop-the-other, calculation of interlayer interactions contribution to the total anisotropy energy can be reduced to the one-dimensional problem summing terms like $$U^{\alpha\beta}=\sum_{n=1}^N S_n^\alpha S_{n+1}^\beta$$ Taking parametrization \eqref{eqn:spiral}, which includes both incommensurate and commensurate structures, one obtains
\begin{eqnarray}
U^{\alpha\beta}&=&\sum_n \frac{1}{2}l_1^\alpha l_1^\beta \left(\cos qa+\cos(2 q x_n+q a)\right)+\nonumber\\
&&+\frac{1}{2}l_2^\alpha l_2^\beta \left(\cos qa-\cos(2 q x_n+q a)\right)+\nonumber\\
&&\frac{1}{2}l_1^\alpha l_2^\beta \left(\sin qa+\sin(2 q x_n+q a)\right)+\nonumber\\
&&\frac{1}{2}l_2^\alpha l_1^\beta \left(-\sin qa+\sin(2 q x_n+q a)\right)\label{eqn:incom}
\end{eqnarray}
\noindent here $q$ is the structure wave-vector, $a$ is the model chain period and $x_n$ is the $n$-th spin position, $N$ is the number of bonds. For incommensurate structure the phase $(2 q x_n+q a)$ runs over all possible values averaging all corresponding cosine and sine terms  to zero. Hence, per bond:
\begin{eqnarray}
\frac{1}{N} U_\textrm{incomm}^{\alpha\beta}&=&\frac{1}{2}(l_1^\alpha l_1^\beta+l_2^\alpha l_2^\beta) \cos qa+\nonumber\\
&&+\frac{1}{2}(l_1^\alpha l_2^\beta-l_1^\beta l_2^\alpha) \sin q a \label{eqn:anis-incomm}
\end{eqnarray}

Collinear antiferromagnetic structure $qa=\pi$ yields different compact answer
\begin{equation}\label{eqn:anis-pia}
\frac{1}{N} U_{q a=\pi}^{\alpha\beta}=-l_1^\alpha l_1^\beta
\end{equation}

In the simplest case $\alpha=\beta=z$ for incommensurate case $\frac{1}{N} U_\textrm{incomm}^{zz}=\frac{1}{2}(1-(l_3^z)^2) \cos qa$ and for collinear case $\frac{1}{N} U_{qa=\pi }^{zz}=-(l^z)^2$, i.e. inter-layer interactions contribution to the anisotropy energy  changes two-fold. Combined with 2D layer contribution to anisotropy energy the net change of the effective anisotropy constants can be even stronger.
\subsection{Different forms of magnetic ordering in light- and heavy-alkali-ion nabokoites: clues from the finite cluster  toy-model}
\begin{figure}[th]x
\centering
  \includegraphics[width=\columnwidth]{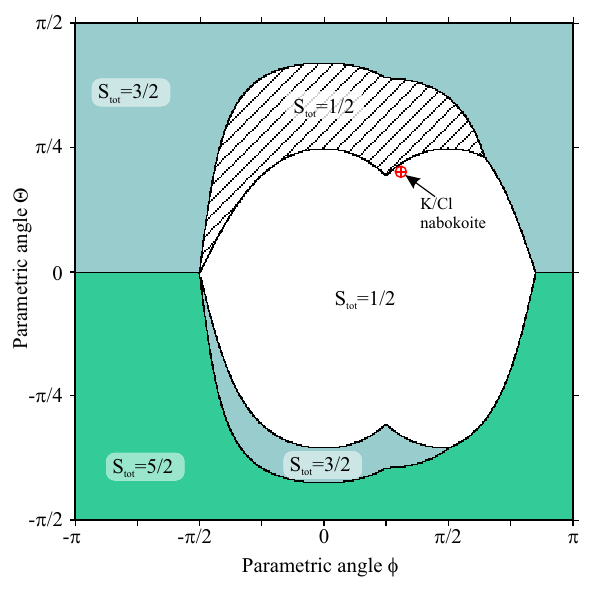}\\
  \caption{(color online) Map of the ground state of the quantum 5-spin pyramid \eqref{eqn:pyramid-ham}, \eqref{eqn:mapping}. Color sectors correspond to different values of total spin of pyramid ground state. $S_\textrm{tot}=1/2$ sector splits into two subsectors: blank white subsector corresponds to the singlet  state $S_\square=0$ at pyramid base, hatched subsector -- to the entangled state of all five spins. Circle symbol marks parameters  of K/Cl-nabokoite according to the  DFT calculations \cite{DFT}.  }\label{fig:quantum-pyramid}
\end{figure}

A qualitative result of our experiments is the strong difference of low-temperature magnetically ordered states in heavy- and light-alkali ion nabokoites. This difference is somewhat surprising in a family of structurally very similar compounds (Fig.~\ref{fig:struct}). Below we discuss a toy-model that provides possible clues for this problem.

Both the nearest-neighbor bonds model (Fig.~\ref{fig:struct}) and the DFT-suggested \cite{DFT} model with next-nearest bonds (Fig.~\ref{fig:j-bonds}) include pyramidal structural block with square base formed by four Cu1 ions and symmetrically positioned apical inter-layer Cu2 ion. Note, that the exact square form of the pyramid base and the equivalence of all Cu1-Cu2 base-apex bonds rely on symmetric high-temperature crystallographic structure, which can be distorted at ferroelectric transition in light-alkali-ion nabokoites.

We consider this pyramidal block as a finite-size cluster and look for the quantum ground state of this problem with spin $S=1/2$ at each pyramid vertex and a Hamiltonian

\begin{eqnarray}
\widehat{\cal{H}}&=&J_2(\widehat{\vec{S}}_1\widehat{\vec{S}}_2+\widehat{\vec{S}}_2\widehat{\vec{S}}_3+\widehat{\vec{S}}_3\widehat{\vec{S}}_4+\widehat{\vec{S}}_4\widehat{\vec{S}}_1)+\nonumber\\
&&+J_4(\widehat{\vec{S}}_1\widehat{\vec{S}}_3+\widehat{\vec{S}}_2\widehat{\vec{S}}_4)+\nonumber\\
&&+J_5(\widehat{\vec{S}}_1+\widehat{\vec{S}}_2+\widehat{\vec{S}}_3+\widehat{\vec{S}}_4)\widehat{\vec{S}}_5\label{eqn:pyramid-ham}
\end{eqnarray}
\noindent here exchange bonds definitions follows \cite{DFT} and Fig.~\ref{fig:j-bonds} and spins enumeration follows Fig.~\ref{fig:j-bonds}.

Since the ground state does not changes at Hamiltonian parameters scaling, we add constraint $$J_2^2+J_4^2+J_5^2=1$$ which allows to map all possible combinations of the bonds relative strength to the spherical surface and to parameterize them as follows:
\begin{eqnarray}
J_5&=&\sin\Theta\nonumber\\
J_2&=&\cos\Theta \cos\varphi\label{eqn:mapping}\\
J_4&=&\cos\Theta \sin\varphi\nonumber
\end{eqnarray}

Now we numerically diagonalize Hamiltonian \eqref{eqn:pyramid-ham} over all parameters space $-\pi\leq\varphi<\pi$, $-\pi/2\leq\Theta\leq\pi/2$ (which includes all possible couplings, both ferro- and antiferromagnetic) and look for the total spin of the cluster ground state, total spin of the pyramid square base and mean projection of the apical spin (in the later case we add weak magnetic field to provide quantization axis and to lift the ground state degeneracy). This analysis was performed via calculation of the quantum averages $\langle S_\textrm{tot}^2 \rangle=\langle\psi_0|(\hat{\vec{S}}_1+\hat{\vec{S}}_2+\hat{\vec{S}}_3+\hat{\vec{S}}_4+\hat{\vec{S}}_5)^2|\psi_0\rangle$,
$\langle S^2_\square \rangle=\langle\psi_0|(\hat{\vec{S}}_1+\hat{\vec{S}}_2+\hat{\vec{S}}_3+\hat{\vec{S}}_4)^2|\psi_0\rangle$ and $\langle S^z_5\rangle=\langle\psi_0|\hat{S}^z_5|\psi_0\rangle$, correspondingly.

The ``world's map'' of this problem is shown at Fig.~\ref{fig:quantum-pyramid}. ``Poles'' $\Theta=\pm\pi/2$ of the sphere correspond to zero coupling within the square base, both along the sides and along the diagonals of the square and $J_5=\pm1$, i.e. a cross-like bonds configuration. At the ``south pole'' $\Theta=-\pi/2$ all couplings are ferromagnetic and the  ground state is fully polarized $S_\textrm{tot}=5/2$ state, at the ``north pole'' $\Theta=\pi/2$ antiferromagnetic couplings on the cross arms lead to $S_\textrm{tot}=3/2$ ground state. ``Equator'' $\Theta=0$ corresponds to the absence of the base-to-apex coupling and apical spin is free along this line. Sector $-\pi\leq\varphi<-\pi/2$ corresponds to the ferromagnetic couplings within the square base, which naturally extends $S_\textrm{tot}=5/2$ and $S_\textrm{tot}=3/2$ sectors up to the equator.  Central part of the map corresponds ro the parameters sector with significant antiferromagnetic couplings, which results in a narrow $S_\textrm{tot}=3/2$ sector in the ``southern hemisphere'' and a central $S_\textrm{tot}=1/2$ area.

Analysis of the pyramid square base spin  and of the apical spin projection revealed that $S_\textrm{tot}=1/2$ total spin sector is split into two subsectors: larger subsector (blank white at Fig.~\ref{fig:quantum-pyramid}) corresponds to the quantum  singlet $S_\square=0$ state of the pyramid base and a totally decoupled apical spin while a smaller subsector (hatched area at Fig.~\ref{fig:quantum-pyramid}) corresponds to the entangled quantum state of all five spins of the pyramid.

Taking the values of exchange couplings predicted by DFT modeling \cite{DFT} we can place K/Cl-nabokoite location on our map at $\varphi=0.968$ and $\Theta=0.630$  (Fig.~\ref{fig:quantum-pyramid}). This location turns out to be very close to the border between the `singlet-at-the-base' and entangled ground state of the pyramidal block. This closeness indicate that a minor change of the exchange coupling parameters could significantly affect the ground state of the pyramidal block and of the whole nabokoite spin system.

For the next-nearest neighbor coupling model \cite{DFT} (Fig.~\ref{fig:j-bonds}) the `singlet-at-the-base' state of the pyramid is effectively decoupled from the  square lattice formed by Cu3-Cu3 $J_6$ bonds. This leaves simple unperturbed square lattice which orders in conventional collinear up-down-up-down pattern once the Ising anisotropy or a weak inter-layer coupling are introduced. If the pyramid ground state is in entangled sector, the Cu3-Cu1 $J_3$ bonds establish frustrating coupling between the square lattice and pyramid total spin. Here the choice of ordered state becomes more complicated  and lies beyond our analysis. Note, that since the change of alkali or halogen ion in \nabok{A}{X}{} affects mostly the inter-layer spacing (Fig.~\ref{fig:struct}), the square lattice remains almost the same --- which could explain closeness of the antiferromagnetic order parameters in heavy-alkali-ion nabokoites if all these compounds fall to the `singlet-at-the-base' sector. On the other hand, distortions of the pyramid parameters bringing it into the entangled sector over the borderline at the map at Fig.~\ref{fig:quantum-pyramid} are individual (especially after the ferroelectric transition) --- which is in agreement with the observed individuality of the ordered state parameters in noncollinearly ordered light-alkali-ion nabokoites.

\section{Conclusions}
Nabokoite family compounds \nabok{A}{X}{} (A=Na, K, Rb, Cs; X=Cl, Br) provide an example of iso-structural  quasy-2D highly frustrated  antiferromagnets with minute changes of lattice parameters. This allow to study how the choice of the ground state in highly-frustrated magnets is affected by fine tuning of interactions balance. While the bulk Curie-Weiss temperatures  $\Theta\simeq 150$~K are practically the same in all nabokoites, the low-temperature ordered states turn out to be qualitatively different in light-alkali-ion (A=Na, K) and heavy-alkali-ion (A=Rb, Cs) nabokoites.

Heavy-alkali-ion nabokoites demonstrate conventional collinear antiferromagnetic ordering of easy-axis type, which is unambiguously evidenced by antiferromagnetic resonance spectroscopy. Characteristics of the antiferromagnetically ordered state (zero-field magnon gaps, see Fig.~\ref{fig:tempres-heavy}, and spin-flop field values, see Table~\ref{tab:temperatures} and Figs.~\ref{fig:heavy-afmr},~\ref{fig:heavy-m}) are almost the same for all four compounds (Rb/Cl, Rb/Br, Cs/Cl and Cs/Br-compounds). This indicates that modification of interlayer spacing via change of halogen or alkali ion does not affect strongly ordered spin system of nabokoite, highlighting that the magnetic ordering takes place within the 2D layers.

Light-alkali-ion nabokoites follow qualitatively different route towards the choice of the low-temperature ground state. Firstly, all light-alkali-ion compounds (Na/Cl, K/Cl and K/Br nabokoites) undergo ferroelectric phase transition (already reported for chlorine series \cite{rebrov,markina,china}) at 25-90~K. Structural adjustment during ferroelectric transition could partially lift the magnetic frustration thus pre-cooking the future magnetic ordering. Secondly, the antiferromagnetic order is formed at low temperatures via two-step phase transition with $T_\textrm{c1}\simeq5-6$~K and $T_\textrm{c2}\simeq 3-4$~K (see Table~\ref{tab:temperatures}). The formed antiferromagnetic order is noncollinear, which is proved by observation of the characteristic non-Larmor antiferromagnetic resonance mode. Electron spin resonance spectroscopy and magnetization measurements demonstrate presence of three zero-field magnon gaps and two critical fields. The lower gap seems to be of anisotropic origin and the critical field related to this mode can be identified as a spin-flop transition, while the larger gaps are of exchange origin and the larger critical field is related to the switching of magnetic structure. Zero-field magnon gaps and critical field values are individual in different light-alkali-ion nabokoites, see Table~\ref{tab:noncoldata}. These observations indicate that light-alkali ion nabokoites are subject to individual balance of some competing interactions. The two-step phase transition is discussed in the context of possible transition between incommensurate and commensurate interlayer ordering patterns.

The possible clue to this qualitative difference of nabokoites properties is provided by finite-size cluster modeling. We consider pyramidal building block of nabokoite structure and solve quantum-mechanical problem \eqref{eqn:pyramid-ham} to find a ground state of this five-spin cluster for different couplings values (Fig.~\ref{fig:quantum-pyramid}). While the values of exchange coupling parameters in nabokoites are not known independently, DFT calculations of \cite{DFT} provides certain estimate, which turns out to be critically close to the border-line between different quantum ground states of the pyramid: one with singlet quantum spin state at the pyramid base and decoupled apical spin and the other with all five spins entangled. Speculatively, case of the heavy-alkali-ion nabokoites corresponds to the singlet state at the pyramid base and ordering within 2D square lattice, while the light-alkali-ion nabokoites features pyramidal blocks with the entangled spin state which brings in frustration to the ordered spin system causing choice of noncollinear magnetic ordering.

While electron spin resonance spectroscopy allows to distinguish different ordering patterns in two subsets of nabokoite family magnets, we cannot for now  decipher exact microscopic ordering pattern in nabokoites. This is a subject of future research with microscopic probes, such as NMR and $\mu$SR, or with neutron diffraction.

\acknowledgements
Authors thank Prof.~A.~Smirnov (Kapitza Institute) for stimulating discussions and Prof. K.A. Lyssenko (Lomonosov Moscow State University) for help with cell constants determination on single crystals for some compounds. Authors acknowledge usage of Balls\&Sticks software for visualization of crystal structure images \cite{bs}.

The work at P.~Kapitza Institute for Physical Problems (magnetic resonance experiments and data analysis) was supported by RSF~22-12-00259-$\Pi$. Part of the research (dielectric permittivity measurements) was conducted under the state assignment of Lomonosov Moscow State University.  Some of the authors (MMM and ANV) acknowledges the support by the Ministry of Science and Higher Education of the Russian Federation in the framework of the Strategic Academic Leadership program ``Priority 2030'' (MISIS Strategic Technology Project `Quantum Internet').

\appendix

\section{Powder averaged antiferromagnetic resonance  absorption for collinear antiferromagnets\label{sec:app-pow}}
\begin{figure*}[th]
\centering
  \includegraphics[width=\textwidth]{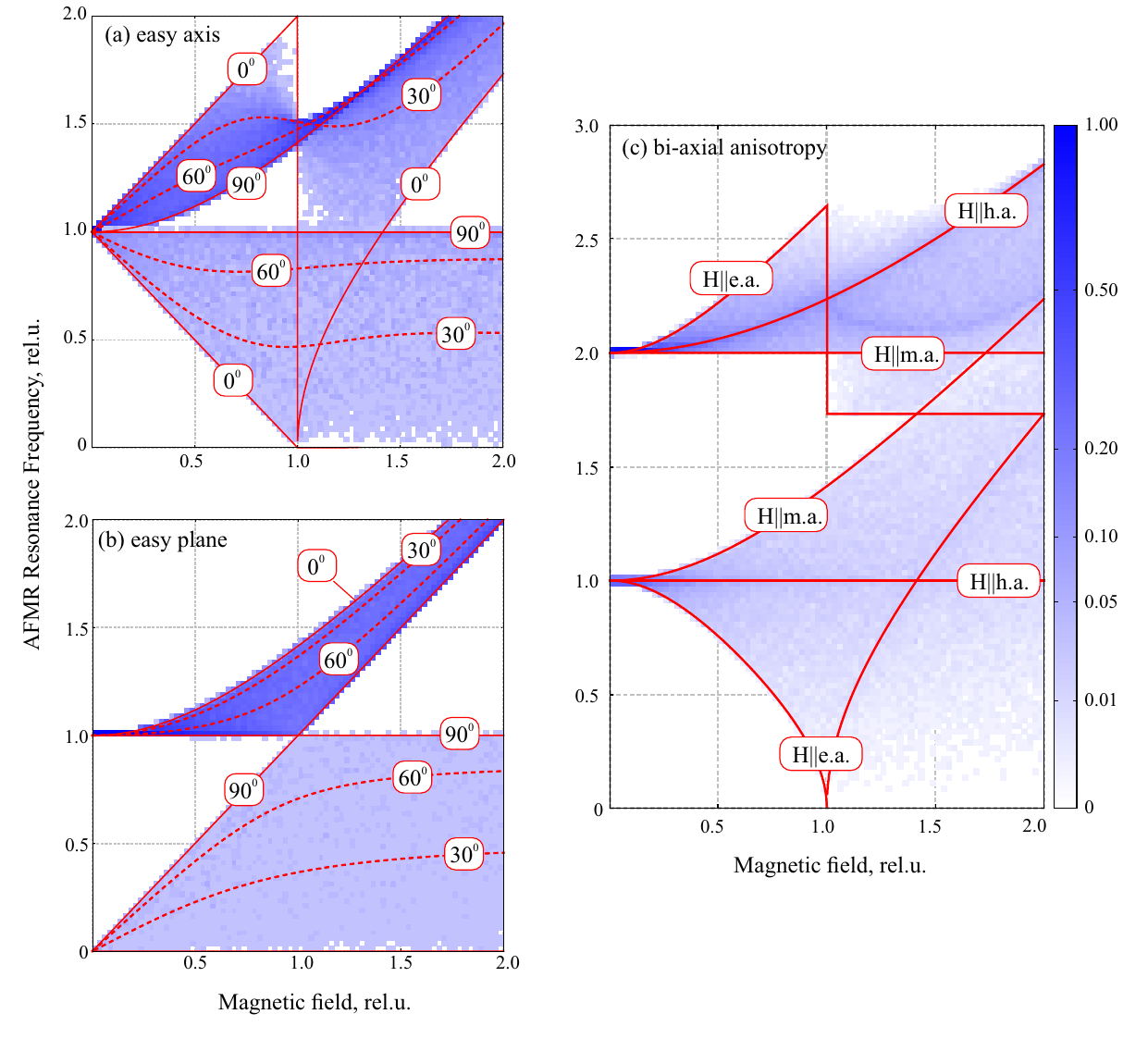}\\
  \caption{(color online) Distribution of the antiferromagnetic resonance (AFMR) eigenfrequencies for the powder sample of collinear antiferromagnet. (a) Case of the easy-axis anisotropy, (b) Case of easy-plane anisotropy, (c) Case of bi-axial anisotropy. Color map shows probability density distribution, color-bar scale is the same for all panels. Curves show $f(H)$ dependences for the particular field orientation: field orientation is given as an angle from the main anisotropy axis for (a) and (b) panels, ``e.a'', ``m.a.'' and ``h.a.'' at the (c) panel stay for easy, middle and hard anisotropy axis correspondingly.  }\label{fig:col-pow-model}
\end{figure*}

Antiferromagnetic resonance is a $q=0$ oscillations of the ordered spin structure. In the case of collinear antiferromagnet in the relatively weak fields $\mu_\textrm{B} B\ll J$  order parameter at $T\approx 0$ can be described by unit vector $\vec{l}$, which is the normed sublattice magnetization for the conventional two-sublattices antiferromagnet (for the model 1D antiferromagnetically ordered chain $\vec{S}_i=(-1)^i \vec{l}$). Vector field $\vec{l}(\vec r, t)$ describes distribution of the order parameter over the crystal and its evolution in time domain. Lagrangian density for this vector field is  \cite{andmar}:

\begin{equation}
\label{eqn:col-L}
{\cal L}=\frac{\chi_\perp}{2\gamma^2} \left({\dot{\vec{l}}}+\gamma[\vec{l}\times  \vec{H}]\right)^2-\frac{a_1^2}{2} l_x^2-\frac{a_2^2}{2} l_y^2
\end{equation}

\noindent
here $\chi_\perp$ is the transverse susceptibility of the collinear antiferromagnet (longitudinal susceptibility is zero at $T=0$). Preferable order parameter orientation is determined by the last two terms:  $a_1=a_2=a>0$ corresponds to the easy-axis anisotropy case with $z$-axis being the easy axis (at zero field $\vec{l}\parallel z$); $a_1=a_2=a<0$ corresponds to the easy-plane anisotropy case with $(xy)$ plane being the easy plane (at zero field $\vec{l}\perp z$); $a_2>a_1>0$ corresponds to the bi-axial case with $z$-axis being the easy axis, $x$-axis --- the middle axis and $y$-axis --- the hard axis. Characteristic spin-flop transition appears as the field is applied along the easy axis at $H_\textrm{SF}=\sqrt{a_1/\chi_\perp}$: at this field order parameter rotates away from the easy axis as the loss in anisotropy energy is overcome by the gain in magnetization energy.

The dynamic equations can be derived via Lagrange-Euler equations yielding two eigenmodes of order parameter oscillations,  the resulting $f(H)$ dependences are the same as for the two-sublattice model \cite{kubo,gurevich}. Characteristic features of the $f(H)$ dependences are the zero-field spin-waves gaps and softening of one of the AFMR modes at spin-flop transition. For the easy-axis case both modes are gapped with the same gap $\Delta=\gamma\sqrt{a/\chi_\perp}$, for the easy-plane case only one mode is gapped $\Delta=\gamma\sqrt{\abs{a}/\chi_\perp}$ and the other is gapless, for the bi-axial case both modes are gapped with different gaps $\Delta_{1,2}=\gamma\sqrt{a_{1,2}/\chi_\perp}$.

We are dealing with the powder sample and AFMR absorption have to be averaged over powder particles orientations. To simplify this problem we neglect details of the dynamic susceptibility $\chi''(\omega,H)$ field and polarization dependence and calculate distribution of resonance eigenfrequencies on $(f, H)$ plane for a magnetic field of random magnitude with directions equally distributed over the sphere. For the modeling we took $\chi=1$, $\gamma=1$, $a_1=a_2=\pm1$ for the easy-axis and easy-plane case, $a_1=1$ and $a_2=4$ for the bi-axial case. AFMR eigenfrequencies for the randomly chosen field were calculated using Octave script \cite{numa-scripts}. Resulting distributions calculated for 100,000 field vector values are shown at Fig.~\ref{fig:col-pow-model}.

Note that resulting figures are qualitatively different for different kinds of anisotropy. Bi-axial anisotropy is distinct by the presence of two gaps. Easy-plane anisotropy yields absorption bands at $f>\Delta$ with right edge at paramagnetic resonance position and very broad (and hence of low absorption amplitude) absorption band at $f<\Delta$ with left edge at the paramagnetic resonance position. Easy-axis anisotropy yields intense absorption band at $f>\Delta$ located to the left of paramagnetic resonance position and weak absorption band at $f\gtrsim\Delta$ close to spin-flop field, at lower frequencies $f<\Delta$ left edge of absorption for easy-axis case shifts linearly as $\gamma H_\textrm{left}=\Delta-f$.

For the sake of the present paper, comparison of the modeling results (Fig.~\ref{fig:col-pow-model}) with the experimental data (Fig.~\ref{fig:heavy-afmr}) clearly reveals that only easy-axis anisotropy case qualitatively fits to the observed AFMR pattern.

\section{Smearing of the spin-flop transition in powdered collinear  easy-axis antiferromagnet \label{sec:app-sf}}
\begin{figure}[th]
\centering
  \includegraphics[width=\columnwidth]{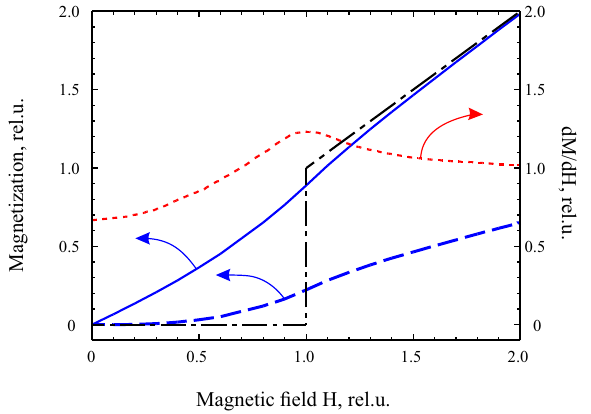}
  \caption{(color online) Solid curve: magnetization process of the powdered easy-axis antiferromagnet.  Model parameters (see Eqn.~\eqref{eqn:Pot-col}) $\chi_\perp=1$, $a=1$. Dash-dotted curve: magnetization curve for the field applied along the easy axis. Dashed curve: magnetization curve for powdered easy axis antiferromagnet with subtracted low-field linear dependence $\Delta M=M(H)-\frac{2}{3} \chi_\perp H$. Dotted curve: differential susceptibility $dM/dH$ for the powdered easy-axis antiferromagnet. }\label{fig:Pow-M-model}
\end{figure}

Static magnetization of the collinear easy-axis antiferromagnet demonstrates  characteristic jump at a spin-flop transition once the gain in the magnetization energy overcomes the loss in anisotropy energy. Sharp spin-flop transition is observed only for the field applied along the easy axis. Powder averaging smears this anomaly and it turns instructive to model magnetization curve of powdered easy axis antiferromagnet to compare modeling results with the experiment.

Static magnetization of a single particle can be modeled using the same Lagrangian~\eqref{eqn:col-L}. Potential energy for the given powder particle is

\begin{equation}
\label{eqn:Pot-col}
\Pi=-\frac{\chi_\perp}{2} \left[\vec{l} \times \vec{H}\right]^2+\frac{a}{2}\left(l_x^2+l_y^2\right)
\end{equation}

\noindent this potential energy have to be minimized to find equilibrium orientation of the order parameter $\vec{l}_0$. Magnetization can be found as $\vec{M}=-\frac{\partial \Pi}{\partial \vec{H}}$, however experimentally measured quantity is usually the projection of the magnetization on the field direction

\begin{equation}
\label{eqn:m-proj-col}
M_H=\chi_\perp H \left(1-\left(\vec{l}_0\cdot\vec{n}_H\right)^2\right)
\end{equation}
\noindent here $\vec{n}_H$ is the unit vector in the magnetic field direction.

We performed modeling of $M_H(H)$ curve by taking powder particles with anisotropy axis uniformly distributed over the sphere and averaging their contributions to the net magnetization, model parameters are $\chi_\perp=1$ and $a=1$, which corresponds to $H_\textrm{SF}=1$ for the field applied exactly along the easy axis. Resulting magnetization curve is shown at Fig.~\ref{fig:Pow-M-model}. Powder  magnetization curve
has a non-zero slope even at low fields with $H\rightarrow 0$ limit $M_{H \rightarrow 0}=\frac{2}{3}\chi_\perp H$ and smoothly changes slope practically reaching high-field asymptotic linear behavior limit $M_{H>>H_{SF}}=\chi_\perp H$ at approximately $2 H_\textrm{SF}$. Differential susceptibility $dM/dH$ demonstrates a broad maximum at $H_\textrm{SF}$ (see dotted curve at Fig.~\ref{fig:Pow-M-model}).

When antiferromagnetic response coexists with some linear magnetization process, spin-flop transition in powdered sample can be better visualized by subtracting low-field linear magnetization law (see dashed curve at Fig.~\ref{fig:Pow-M-model}). Resulting $\Delta M(H)$ curve demonstrates smooth increase close to $H_\textrm{SF}$, its value at transition field

 \begin{equation}
 \label{eqn:dM(Hsf)}
 \Delta M(H_\textrm{SF})\approx 0.22 \chi_\perp H_\textrm{SF}.
 \end{equation}

\section{AFMR frequencies for the noncollinear antiferromagnet without anisotropy \label{sec:app-noncol}}
Following \cite{andmar} low-energy spin dynamics of noncollinear antiferromagnet (e.g., spiral or non-coplanar spin structure) can be parameterized by the oscillations of three mutually orthogonal unit vectors $\vec{l}_{1,2,3}$ with the Lagrangian density \cite{svistfar}

\begin{eqnarray}
{\cal L}&=&\frac{I_1}{2}\left({\dot{\vec{l}}}_1+\gamma\left[{\vec{l}}_1\times \vec {H}\right]\right)^2+\frac{I_2}{2}\left({\dot{\vec{l}}}_2+\gamma\left[{\vec{l}}_2\times \vec {H}\right]\right)^2+\nonumber\\
&+&\frac{I_3}{2}\left({\dot{\vec{l}}}_3+\gamma\left[{\vec{l}}_3\times \vec {H}\right]\right)^2-U_\textrm{anis}({\vec{l}}_1, {\vec{l}}_2, {\vec{l}}_3) \label{eqn:noncol-L}
\end{eqnarray}

\noindent here $U_\textrm{anis}({\vec{l}}_1,{\vec{l}}_2{\vec{l}}_3)$ is the anisotropy energy depending on orientation of antiferromagnetic vectors with respect to the crystal axes. Neglecting anisotropy effects, parameters $I_i$ can be expressed via static susceptibilities. Defining $\chi_i$ as the susceptibility for the magnetic field applied along the ${\vec{l}}_i$, one obtains

\begin{eqnarray}
\chi_1&=&\gamma^2 (I_2+I_3)\nonumber\\
\chi_2&=&\gamma^2(I_1+I_3)\label{eqn:chi-I}\\
\chi_3&=&\gamma^2(I_1+I_2)\nonumber
\end{eqnarray}

Eigenfrequencies of the order parameter oscillations can be found via Euler-Lagrange equations which yields three low-frequency AFMR modes for noncollinear case. In the absence of anisotropy energy  ($U_\textrm{anis}=0$) under assumption $\chi_3>\chi_2>\chi_1$ ($I_1>I_2>I_3$) equilibrium state corresponds to ${\vec{l}}_3\parallel \vec {H}$ and

\begin{eqnarray}
\omega_1&=&\gamma H\nonumber\\
\omega_2&=&\gamma H \sqrt{\frac{(I_1-I_3)(I_2-I_3)}{(I_1+I_3)(I_2+I_3)}}=\nonumber\\
&=&\gamma H\sqrt{\frac{(\chi_3-\chi_1)(\chi_3-\chi_2)}{\chi_1 \chi_2}}\label{eqn:omega-no-anis}\\
\omega_3&=&0\nonumber
\end{eqnarray}

\noindent Zero frequency $\omega_3$ mode corresponds to the degenerate rotations around the field direction. Mode $\omega_1$ is the conventional Larmor mode, while $\omega_2$ is the non-Larmor mode. Its anomalous slope factor (see Eq.~\eqref{eqn:gamma-factor}) is determined by susceptibilities ratio

\begin{equation}\label{eqn:app-gamma-chi}
\Gamma=\sqrt{\frac{(\chi_3-\chi_1)(\chi_3-\chi_2)}{\chi_1 \chi_2}}
\end{equation}
\noindent and can be both above and below unity.

It is worthy to note, that while anisotropic spin-spin interactions are not required to provide anomalous non-Larmor eigenfrequency, the dynamic susceptibility of this mode is zero in the exactly Heisenberg case. This follows from fluctuation-dissipation theorem or van Vleck's theory of moments  \cite{altkoz} which proves that symmetric Heisenberg exchange interaction alone does not shift resonance absorption from the Larmor frequency. Anisotropic spin-spin interactions provide non-zero dynamic susceptibility for this mode and makes it visible in magnetic resonance experiment.

\end{document}